# Multilevel genomic constraints shape nuclear tRNA gene organization in plants

Guillaume Hummel[a, #], David Pflieger[a, #], Valérie Cognat[a], Laurence Drouard[a]*, Alexandre Berr[a]*

[a]Institut de biologie moléculaire des plantes, CNRS, Université de Strasbourg, 12 rue du Général Zimmer, F-67084 Strasbourg, France.

[#]These authors contributed equally

*Corresponding authors: Laurence Drouard and Alexandre Berr
Emails: Laurence.drouard@cnrs.fr / Alexandre.berr@cnrs.fr

## SUMMARY (max 250 words / actual version = 249)

Transfer RNAs (tRNAs) are essential components of the translational machinery. Their abundance and diversity shape decoding capacity and protein synthesis efficiency and accuracy. Because tRNA abundance is encoded in the genome through tDNA copy number, chromosomal organization, and *cis*-regulatory sequences controlling transcription, these features are expected to influence translational. However, the principles governing nuclear tDNA organization remain poorly understood. Here, we analyzed nuclear tDNA repertoires across 53 photosynthetic eukaryotes spanning major Archaeplastida lineages and secondary endosymbionts, along with seven non-plant eukaryotic outgroups, using comparative genomic approaches at sequence, chromosomal, and genome-wide scales. To support these analyses and enable interactive exploration of tDNA organization, we developed ShinytRNA (https://nebula.ibmp.unistra.fr/shinytRNA/), a web application for genome-wide exploration of tDNA organization. Nuclear tDNA copy numbers vary by more than two orders of magnitude across species, yet the relative representation of tRNA families corresponding to each amino acid remains strikingly conserved across lineages, revealing strong evolutionary constraints on tDNA dosage. In angiosperms, tDNAs show reinforced cis-regulatory features linked to RNA polymerase III transcription, including expanded AT-rich regions, enriched CAA motifs, and extended poly(T) tracts. At the chromosomal scale, tDNAs are predominantly dispersed along chromosome arms, with homogeneous spacing that scales with genome size, while also showing non-random chromosomal distribution, exclusion from centromeric regions, and limited clustering. Together, these patterns reveal conserved yet lineage-specific principles governing nuclear tDNA organization in plants, and highlight how multiple genomic constraints shape the evolution of nuclear tDNA repertoires.

## INTRODUCTION

Transfer RNAs (tRNAs) are essential components of the translational machinery that decode messenger RNA (mRNA) codons into amino acids during protein synthesis. Each tRNA molecule carries a specific amino acid and recognizes one or several synonymous codons through its anticodon, thereby ensuring accurate and efficient translation (Rodnina and Wintermeyer, 2001). The abundance and diversity of cellular tRNA pools are key determinants of translational efficiency and fidelity (Plotkin and Kudla, 2011) (Novoa and Ribas de Pouplana, 2012) (Zhou *et al.*, 2013) (Kirchner and Ignatova, 2015).

Part of the regulation and efficiency of protein translation is encoded in genomes through the composition and organization of nuclear tRNA gene (tDNA) repertoires. Nuclear tDNAs form multi-copy families whose abundance varies widely across species and lineages (Goodenbour and Pan, 2006, Chan and Lowe, 2016) (Santos and Del-Bem, 2022). Because gene copy number often correlates with relative tRNA abundance, tDNA repertoires are commonly used as a proxy for decoding capacity across tRNA isotypes (*i.e.*, tRNAs carrying the same amino acid) (Duret, 2000) (Michaud *et al.*, 2011) (Novoa and Ribas de Pouplana, 2012). However, copy number alone provides only a partial view of functional tRNA supply, as individual loci can differ markedly in transcriptional activity, and subsets of tDNAs may be selectively deployed across tissues and developmental programs (Sagi *et al.*, 2016), (Torres *et al.*, 2019),(Gerber *et al.*, 2020) (Hummel *et al.*, 2020) (Thornlow *et al.*, 2020) (Gao *et al.*, 2024) (Reimao-Pinto *et al.*, 2024) (Hummel *et al.*, 2025).

Transcription of nuclear tDNAs is carried out by RNA polymerase III (Pol III) through a promoter architecture combining conserved internal promoter elements with additional regulatory signals located in the surrounding genomic context (Berg and Brandl, 2021) (Kessler and Maraia, 2021). Two internal promoter elements, the A- and B-boxes, recruit transcription factor TFIIIC, which in turn positions TFIIIB upstream to enable Pol III initiation. In addition to these internal elements, upstream and downstream sequences contribute to transcriptional efficiency, including upstream AT-rich, CAA motifs, and downstream poly(T) termination signals, which collectively modulate transcriptional initiation, termination, and Pol III recycling (Ulmasov and Folk, 1995) (Dieci and Sentenac, 1996) (Choisne *et al.*, 1998) (Yukawa *et al.*, 2000), (Yukawa *et al.*, 2011), (Giuliodori *et al.*, 2003) (Zhang *et al.*, 2011) (Girbig *et al.*, 2022) (Xie *et al.*, 2022) (Monloy and Planta, 2024), (Berg and Brandl, 2021).

Beyond local regulatory features, the chromosomal organization of tDNAs may also influence their functional deployment. Nuclear tDNAs can occur as isolated loci or be organized into clusters, sometimes arising through tandem duplication events (Bermudez-Santana *et al.*, 2010) (Hummel and Liu, 2022) (Monloy and Planta, 2024) (Zhang *et al.*, 2025). In *Arabidopsis thaliana*, clustered tDNAs are epigenetically silenced, illustrating how chromosomal context and chromatin environment can determine which members of the tDNA repertoire remain transcriptionally accessible (Hummel *et al.*, 2020, Hummel and Liu, 2022).

Together, these observations indicate that nuclear tDNA organization involves multiple genomic features, including gene copy number, *cis*-regulatory sequences associated with Pol III transcription, and large-scale chromosomal distribution. However, these features have often been examined separately and in a limited number of model species, leaving unclear whether common organizational patterns emerge across plant genomes and photosynthetic lineages. A broader comparative framework is therefore needed to identify conserved and lineage-specific principles of nuclear tDNA organization from an evolutionary and genomic perspective.

Here, we address this gap by assembling a curated comparative dataset of nuclear tDNA annotations across a broad phylogenetic sampling of photosynthetic eukaryotes together with representative non-plant outgroups (Figure 1a). To enable standardized chromosome-scale exploration of nuclear tDNA organization, we also developed ShinytRNA, a freely accessible interactive web application designed to visualize and export quantitative metrics of tDNA genomic organization (https://nebula.ibmp.unistra.fr/shinytRNA/); (Figure 1b, Figure S1). Developed as an extension of the PlantRNA database (http://plantrna.ibmp.cnrs.fr/), (Cognat *et al.*, 2022), ShinytRNA provides an unified framework for comparative analyses of nuclear tDNA repertoires across genomes. Using the curated comparative dataset and the ShinytRNA framework, we show that nuclear tDNA organization follows conserved yet lineage-specific principles across photosynthetic eukaryotes. Despite large variation in total tDNA number, the relative abundance of tRNA isotype families remains remarkably conserved, revealing strong constraints on tDNA dosage. Angiosperms display coordinated reinforcement of Pol III-associated *cis*-regulatory elements, and land plants maintain a predominantly dispersed chromosomal organization characterized by homogeneous intergenic spacing, exclusion from centromeric regions, and limited clustering restricted to specific isotypes. Together, these patterns reveal how multiple genomic constraints collectively shape the evolution and organization of nuclear tDNA repertoires across plant genomes.

## RESULTS

### Conserved hierarchy of tDNA copy numbers among tRNA families across eukaryotes

Nuclear tDNA copy numbers vary by more than two orders of magnitude among eukaryotic species (Hummel *et al.*, 2019) (Chan and Lowe, 2016). Whether this variation alters the relative abundance of individual tRNA isotypes or instead preserves a conserved hierarchy across lineages remains unclear. tDNA copy number is known to correlate strongly with measured tRNA abundance in yeast and animals (Behrens *et al.*, 2021) (Nagai *et al.*, 2021) (Duret, 2000). In photosynthetic organisms, earlier studies reported correlations between tDNA copy numbers, amino acid usage, and codon frequencies, yet these analyses were limited to only four angiosperms (*i.e.*, *Medicago truncatula*, *Populus trichocarpa*, *A. thaliana*, and *Oryza sativa*) and the green alga *Chlamydomonas reinhardtii* (Michaud *et al.*, 2011). Whether such a relationship is conserved across diverse photosynthetic eukaryotic lineages remained unknown. To address this, we quantified nuclear tDNA copy numbers per isotype across representative photosynthetic lineages, using species with curated tDNA annotations spanning aquatic taxa (*i.e.*, glaucophytes, rhodophytes, chlorophytes, streptophyte algae, and secondary endosymbionts) to land plants (*i.e.*, hornworts, liverworts, mosses, lycophytes, ferns, gymnosperms, and angiosperms), (Figure 1a). To avoid biases associated with local tandem amplification, clustered tDNAs were excluded from this analysis. This choice is further supported by their epigenetic silencing in *A. thaliana* (Hummel *et al.*, 2020).

Despite extreme variation in total tDNA number, ranging from only 43 genes in the highly compact genome of the red alga *Cyanidioschyzon merolae* to approximately 7,500 genes in the fern *Azolla filiculoides* (Cognat et al., 2022), the relative ranking of tDNA copy numbers per isotype remains remarkably conserved (tDNA Isotype module of the ShinytRNA platform; Figure 2). Isotypes encoding rare amino acids, such as tryptophan, methionine, histidine, and cysteine, consistently rank among the least abundant, whereas those for leucine, glycine, alanine, arginine, and serine rank among the most abundant. This hierarchy is conserved not only across deeply diverged photosynthetic taxa but also extends to non-photosynthetic eukaryotic supergroups, including *Excavata*, *Amoebozoa*, and *Opisthokonta* (Figures 1a and 2). Together, these findings reveal a broadly conserved tDNA copy number hierarchy across eukaryotes, pointing to strong and universal translational constraints on tDNA dosage.

**Coordinated reinforcement of Pol III *cis*-regulatory elements in angiosperms**

Beyond gene dosage, the functional contribution of nuclear tDNAs also depends on the *cis*-regulatory elements that control their transcription. To investigate whether the *cis*-regulatory architecture of nuclear tDNAs has diversified across photosynthetic lineages, we analyzed the 5′ upstream (−50 to −1) and 3′ downstream (+1 to +25) flanking regions of nuclear tDNAs (Figure 3). We focused on motifs critical for Pol III transcription, including upstream CAA motifs, AT-rich regions, and poly(T) termination signals. Sequences were retrieved from the PlantRNA 2.0 database (Cognat *et al.*, 2022) and analyzed by sequence logos together with quantitative motif profiling across all annotated nuclear tDNA loci.

Upstream *cis*-regulatory elements display marked lineage-specific patterns across photosynthetic eukaryotes. Approximately 67% of nuclear tDNAs in angiosperms harbor at least one CAA motif within the -10 to -3 upstream region, compared to substantially lower proportions in other photosynthetic lineages and non-plant organisms (Figures 3b and 3c; Figures S2 and S3; Table S2). Notably, CAA motifs are already detectable in the earliest-diverging angiosperms *Amborella trichopoda* (Amborella Genome, 2013), although at lower frequency (48%) than in derived angiosperms (Figures S2 and S3). In addition to this increased occurrence, the position of the first detected CAA motif is more tightly constrained in angiosperms, indicating enhanced positional specificity. Similarly, upstream AT-rich regions between positions -35 and −20 show significantly higher enrichment in angiosperms than in non-angiosperm embryophytes, other Archaeplastida, or lineages derived from secondary endosymbionts (Figures 3b and 3c; Figure S4; Table S2). In contrast, non-angiosperm photosynthetic lineages display lower frequencies of CAA motifs and weaker, more variable AT enrichment profiles (Figure 3c). Downstream termination features are similarly reinforced in angiosperms. Both the overall thymidine content within the +1 to +25 region and the length of the longest contiguous poly(T) stretch are significantly increased compared to other lineages (Figures 3b and 3c; Figures S5-S6; Table S2).

Together, these results reveal a coordinated reinforcement of multiple *cis*-regulatory elements associated with Pol III transcription in angiosperms. The coincident enhancement of upstream AT-rich regions, CAA motifs, and poly(T) termination signals suggests a global optimization of transcriptional initiation, termination, and reinitiation efficiency in flowering plant tDNAs. Notably, although these *cis*-regulatory features are detectable in non-photosynthetic eukaryotic species, they do not exhibit the coordinated enrichment observed in angiosperms (Figures 3b and 3c; Figures S2-S6), indicating that this reinforcement is not a general property of eukaryotic tDNAs but is instead specific to photosynthetic lineages, particularly angiosperms.

**Intergenic spacing of nuclear tDNAs scales with genome size**

Like any multi-copy gene family, the functional output of nuclear tDNAs may also be influenced by their organization, including their spacing and distribution along chromosomes. Whether such chromosome-scale organization follows conserved principles across plant genomes remains largely unexplored. To address this question, we analyzed curated tDNA annotations from 31 photosynthetic species with chromosome-level genome assemblies, retrieved from the PlantRNA 2.0 database and examined using the interactive ShinytRNA platform (tDNA Distribution and tDNA Pairwise modules, Figure 1b). These species span major plant lineages, including one alga, one bryophyte, one lycophyte, one gymnosperm, and 27 angiosperms. For comparison, we also included six representative non-plant outgroups, comprising one protist, one fungus, and four metazoans.

In *A. thaliana*, nuclear tDNAs are distributed across all chromosomes and display relatively uniform inter-tDNA distances (*d*), defined as the genomic distance between adjacent tDNA loci (Figure 4a). Chromosomes 1 and 2, however, exhibit reduced median *d* values and distributions skewed toward smaller distances due to the presence of previously described tDNA clusters (Michaud *et al.*, 2011) (Hummel *et al.*, 2020). When these clustered loci are excluded, median *d* values become comparable across chromosomes, restoring a homogeneous genome-wide spacing pattern (Figure 4a). Thus, in *A. thaliana*, local clustering increases tDNA density without disrupting overall chromosomal dispersion.

This predominantly homogeneous spacing pattern is conserved across angiosperms and, more broadly, among land plants (Figures S7 and S8). In all examined species, chromosomes harboring tDNA clusters display reduced median *d* values relative to genome-wide averages, reflecting local compression caused by densely packed loci (Figure 4b, red boxes; Figure S7). For example, this can be observed in *Zea mays*, *Vitis vinifera* or *Malus domestica* which harbor $tDNA^{Ala}$, $tDNA^{Ile}$, $tDNA^{Pro}$, $tDNA^{Ser}$ clusters, respectively (Table S3; Figure S7). Exclusion of intra-cluster intervals reduces variance and restores more homogeneous spacing distributions, indicating that short within-cluster distances account for most deviations from uniformity (Figure 4b).

To assess how genome expansion relates to tDNA organization, we examined the relationship between genome size, total tDNA number, and median *d* (Figure 4c). Across photosynthetic eukaryotes, total tDNA copy number shows no significant correlation with genome size ($R^2 = 0.13$), suggesting that increases in genome size are not associated with proportional expansion of tDNA repertoires. In contrast, genome size is positively correlated with median *d* ($R^2 = 0.61$), indicating that larger genomes tend to exhibit increased intergenic spacing between tDNA loci. These results suggest that genome expansion is primarily accommodated through increased spacing rather than extensive amplification of tDNA copy number. Notably, the green alga *C. reinhardtii* deviates from this pattern, displaying irregular inter-tDNA distances and multiple dense clusters (Figures S8 and S9; (Cognat *et al.*, 2008)). This organization contrasts with the predominantly homogeneous dispersion observed in land plants and more closely resembles the distribution patterns described in metazoan genomes such as human and mouse (Van Bortle *et al.*, 2017) (Raab *et al.*, 2012) (Iben and Maraia, 2014) (Figure S8).

Collectively, these analyses reveal that land plant genomes exhibit a conserved, predominantly dispersed chromosomal organization of nuclear tDNAs. Although local clustering can generate regional increases in tDNA density, genome expansion is primarily associated with increased spacing between tDNAs rather than their total number.

**Chromosomal partitioning of tDNAs reveals isotype-specific biases and centromeric exclusion**

Beyond intergenic spacing, we next assessed whether tDNA isotype families are randomly distributed across chromosomes relative to genome-wide expectations. Heatmap analyses reveal chromosome-specific enrichment or depletion of particular isotype families relative to expected counts normalized to chromosome size (Figure 4d; Figure S10; tDNA isotype distribution module of the ShinytRNA platform). In *A. thaliana*, for example, tDNA$^{Leu}$ and tDNA$^{Phe}$ show strong enrichment on chromosome 5, whereas tDNA$^{Asp}$ is relatively depleted from chromosome 4. Similarly, tDNA$^{Ile}$ and tDNA$^{Glu}$ copies are overrepresented on chromosome 5 of *Marchantia polymorpha,* whereas tDNA$^{Gln}$ copies show enrichment on chromosome 21 of *Nicotiana tabacum* (Figure 4d). Comparable chromosome-specific enrichments and depletions are also observed in *Oryza sativa*, *Selaginella kraussiana*, and *Ginkgo biloba* (Figure S10), indicating that dispersed tDNAs are not uniformly partitioned among chromosomes from diverse plant lineages.

At a broader scale, tDNA loci preferentially localize along chromosome arms and are markedly depleted from centromeric and pericentromeric regions (Figures S11; Centromere positioning module of the ShinytRNA platform). This exclusion pattern is consistently observed despite major differences in centromere size and organization across species, for example in *A. thaliana*, where centromeres are relatively compact and well defined, as well as in *Solanum lycopersicum*, with much larger chromosomes and more extended pericentromeric regions (Figure S11).

Although clustering influences intergenic spacing at a local scale, these chromosome-level patterns indicate that nuclear tDNAs are also shaped by higher-order genomic compartmentalization. Together, these analyses show that nuclear tDNAs in land plants display non-random chromosomal partitioning, characterized by isotype-specific biases and preferential localization outside centromeric domains. This organization suggests that tDNA loci are integrated into large-scale chromosomal architecture, reflecting conserved spatial constraints that extend beyond simple gene density or local clustering effects.

**tDNA local clustering reveals lineage- and family-specific aggregation patterns**

Although nuclear tDNAs are predominantly dispersed along chromosomes, they also form localized clusters within chromosomal arms. To characterize these clusters, we analyzed their occurrence using the tDNA Clusters module of the ShinytRNA platform across a range of proximity-based criteria applied to chromosome-scale genome assemblies. Clusters were defined as groups of ≥3, ≥5, or ≥10 tDNAs located within genomic windows of 1, 5, or 50 kb (Figure 5; Figure S12). Stringent criteria (≥3 genes within 1 kb) identified only a limited number of loci, whereas more permissive thresholds (≥3 genes within 50 kb) generated numerous associations. Based on this systematic evaluation, we retained a threshold of ≥5 genes within 5 kb as a balanced criterion capturing recurrent clusters while limiting spurious associations (Figure 5). Notably, the use of more permissive criteria (≥10 genes within 50 kb) did not substantially increase clustering ratios in gymnosperms and angiosperms, indicating that plant tDNA clusters are typically compact and tightly bounded. By contrast, in several metazoan genomes (*e.g.*, human and mouse) and in non-plant eukaryotes such as *Saccharomyces cerevisiae* and the protist *Dictyostelium discoideum*, the more permissive criterion (≥10 genes within 50 kb) markedly increased the detectable clustering fraction (Figure S12). This reflects the presence of larger and more spatially extended aggregation domains involving multiple tDNA isotype families, commonly referred to as tDNA islands (Coughlin *et al.*, 2009, Iben and Maraia, 2014), which clearly differ from the compact and isotype-restricted clusters observed in land plants.

Under the threshold, 5 genes within 5kb, the proportion of clustered tDNAs varies markedly across plant species (Figure 5). In most land plants, clustered loci represent a small fraction of total tDNAs, typically below 5%, consistent with a predominantly dispersed genomic organization. In contrast, several *Brassicaceae* species display substantially higher clustering ratios, although this pattern is not universal, as illustrated by *Eutrema parvula*. Outside this lineage, elevated clustering is also observed in *Z. mays*, where clustered tDNAs exceed 10% of total loci. Finally, in the green alga, *C. reinhardtii*, more than 30% of tDNAs are clusterized.

To further characterize the composition of tDNA clusters, we calculated for each isotype a clustering ratio, defined as the proportion of genes belonging to clusters relative to the total gene count of that isotype. This analysis confirms that clustering is unevenly distributed across both lineages and isotype families (Figure 5; Figures S12 and S13). In most photosynthetic protists, including *O. tauri*, *C. merolae*, and *P. tricornutum*, clustering is minimal, with nuclear tDNAs occurring predominantly as singletons. In contrast, land plants display detectable clustering involving a restricted subset of isotypes. Among angiosperms, clustering is dominated by a limited number of recurrent configurations. In particular, tDNA$^{Pro}$ frequently forms homoclusters composed of adjacent copies of the same isotype, whereas tDNA$^{Ser}$ and tDNA$^{Tyr}$ commonly co-occur in structured heteroclusters (Figure S13 and Table S3). Additional isotypes, such as tDNA$^{Gln}$, also show clustering in specific lineages (*e.g.,* in *Hordeum vulgare* and *Oryza sativa* or in *Fragaria iinumae* and *Rosa chinensis*), while other clusters remain species-specific, as illustrated by tDNA$^{Ala}$ clusters in *Z. mays* or tDNA$^{Ile}$ clusters in Z. mays and *Brassica napus* (Figure 5, Table S3).

Taken together, these results show that tDNA clustering in plants is structured and selective, involving a limited set of isotypes organized into compact and recurrent configurations. In contrast, several metazoan genomes and the green alga *C. reinhardtii* exhibit broader aggregation domains under relaxed proximity thresholds, reflecting more spatially extended and heterogeneous tDNA assemblies (Figures S12 and S13). These lineage-specific differences reveal distinct modes of local tDNA organization superimposed on a globally dispersed chromosomal architecture.

**Clustered and dispersed tDNAs exhibit distinct *cis*-regulatory conservation patterns**

The identification of lineage- and isotype-specific tDNA clusters prompted us to examine whether clustering is associated with distinct *cis*-regulatory sequence patterns beyond simple local duplication. We therefore compared the upstream (−50 bp) and downstream (+25 bp) flanking regions of clustered and dispersed tDNAs within individual isotypes (Figure 6). In *A. thaliana*, clustered tDNAs$^{Pro}$ form tandem arrays in which both upstream and downstream regions are more similar to one another than to dispersed tDNA$^{Pro}$ loci. This increased similarity is particularly evident in downstream regions, where clustered loci display highly homogeneous poly(T) tract organization, whereas dispersed loci show greater variability in termination architecture. By contrast, upstream regions are less strongly conserved, suggesting that constraints on transcription termination may be stronger than those affecting initiation within this cluster. The Ser-Tyr-Tyr (SYY) cluster provides a distinct example of structured conservation. Its organization consists of repeated blocks of an entire SYY unit (Hummel *et al.*, 2020). When tDNAs$^{Tyr}$ are analyzed according to their position within the repeated block, the first and second Tyr copies form distinct groups, indicating preservation of their relative positions during cluster expansion. Both upstream and downstream flanking regions show strong conservation together with pronounced nucleotide compositional biases, supporting a model of segmental duplication followed by selective maintenance of the entire *cis*-regulatory module rather than independent aggregation of individual loci. Similar patterns are observed in monocots. In *Z. mays*, large tandem arrays of tDNAs$^{Ile}$ display highly uniform downstream sequences compared to dispersed copies. In *H. vulgare*, clustered tDNAs$^{Gln}$ likewise show greater *cis*-regulatory similarity within clusters than among dispersed loci.

Together, these observations indicate that tandem duplication likely contributes to cluster formation, but does not fully explain the recurrent conservation of cluster composition and *cis*-regulatory architecture. The recurrence of some cluster configurations, notably Pro-containing clusters and partly SYY-like arrangements, across distantly related angiosperms (Figure 5) suggests that these clusters are unlikely to result solely from recent species-specific duplication events and may instead reflect either recurrent emergence under similar selective constraints or older arrangements maintained across lineages. This broader evolutionary persistence is consistent with the fact that different cluster types display distinct patterns of *cis*-regulatory conservation, likely reflecting different functional constraints. Clusters such as SYY maintain strong conservation of both upstream and downstream regions, suggesting selective preservation of complete transcriptional units, whereas clusters such as Pro show stronger conservation of downstream termination signals, consistent with stronger constraints on transcription termination and Pol III recycling. These results suggest that clustered tDNAs follow evolutionary trajectories distinct from dispersed loci, shaped by both lineage-specific functional constraints and long-term maintenance of specific regulatory architectures.

## DISCUSSION

Despite their essential role in translation and emerging functions in genome organization and chromatin regulation (Raab *et al.*, 2012) (Shukla and Bhargava, 2018) (Iwasaki *et al.*, 2020) (Hamdani *et al.*, 2019) (Guimaraes *et al.*, 2021), the principles governing nuclear tDNA organization across photosynthetic eukaryotes remain poorly understood. In particular, it is unclear whether tDNA repertoires primarily reflect neutral genomic processes or are shaped by selective constraints linked to translational demand and chromatin context. Here, by systematically analyzing tDNA repertoires across diverse photosynthetic lineages using the ShinytRNA platform, we uncover multiple layers of non-random organization operating across genomic scales.

In spite of variations in total nuclear tDNA copy numbers, we found that their relative ranking per isotype remains remarkably conserved across photosynthetic eukaryotes (Figure 2). This hierarchy extends to diverse non-photosynthetic eukaryotic supergroups, including *Excavata*, *Amoebozoa*, and *Opisthokonta* (Du *et al.*, 2017), indicating that it represents a conserved and functionally constrained organization of decoding capacity. Consistent with this, high-copy isotypes correspond to frequently used amino acids, whereas low-copy isotypes for rare amino acids (tryptophan, methionine, histidine, cysteine) may enable regulatory ribosome pausing and co-translational protein folding (Plotkin and Kudla, 2011) (Zhou *et al.*, 2013). Interestingly, this hierarchy shows substantial agreement with Trifonov's consensus chronology of amino acid appearance in the genetic code (Trifonov, 2000), with early-recruited amino acids (glycine, alanine, proline, serine, valine, leucine, glutamic acid) tending to have high tDNA copy numbers. Comparison with the more recent LUCA-based reconstruction (Wehbi *et al.*, 2024) reveals some discrepancies, notably high tDNA abundance for leucine and serine classified as late additions, and low abundance for methionine, cysteine, and histidine classified as early. Nevertheless, the overall pattern supports the idea that tDNA copy numbers may still retain a detectable imprint of primordial proteome composition and early translational constraints. Subsequent lineage-specific evolutionary pressures and translational demands have only moderately reshaped this hierarchy. Together, these observations support a model in which codon usage, amino acid composition, and tDNA dosage remain tightly interconnected.

Despite the strong conservation of the tDNA copy number hierarchy across eukaryotes, total tDNA numbers scale weakly or non-significantly with genome size in photosynthetic lineages (Figure 4); (Michaud *et al.*, 2011), (Monloy and Planta, 2024). This weak relationship contrasts with the positive correlation reported in many animals and fungi (Goodenbour and Pan, 2006) (Bermudez-Santana *et al.*, 2010) (Santos and Del-Bem, 2023) and indicates that tDNA repertoires in photosynthetic eukaryotes are not primarily driven by genome expansion, but by multiple interacting evolutionary processes. Transposable element (TE) proliferation, the main driver of genome expansion in plants, can increase intergenic space without proportional amplification of tDNA copy number (Lee and Kim, 2014), while local tandem duplication may promote selective expansion of specific isotypes (*e.g.* the Pro cluster discussed below; (Zhang *et al.*, 2025)). In addition, whole-genome duplication (WGD), followed by biased gene retention or loss during diploidization, further prevent linear scaling between tDNA copy number and genome size (Panchy *et al.*, 2016). Together, these processes help maintain a dynamic balance between genome expansion, translational efficiency, and the constrained evolution of the tDNA repertoire (Santos and Del-Bem, 2023) (Monloy and Planta, 2024). In this context, modulation of transcriptional activity through *cis*-regulatory elements may represent an important additional layer of regulation.

Consistent with this, our analyses reveal a pronounced enrichment of Pol III-associated *cis*-regulatory elements in angiosperms, whereas other photosynthetic lineages display weaker patterns resembling those of non-plant eukaryotes (Figure 3c). This reinforcement appears to have emerged progressively during angiosperm evolution. The early-diverging angiosperm *A. trichopoda* already exhibits detectable, albeit weaker, enrichment of CAA motifs, indicating that strengthening of *cis*-elements began early in angiosperm diversification and was subsequently amplified in derived lineages. Functionally, these elements are known to modulate multiple steps of Pol III transcription, including initiation, termination, and polymerase recycling (Yukawa *et al.*, 2000). Their coordinated enrichment therefore may likely reflect an optimization of Pol III transcriptional efficiency. It may also signify increased constraints on transcriptional efficiency and regulatory precision associated with genome expansion and organismal complexity in angiosperms (Dieci *et al.*, 2013, Bowman *et al.*, 2017) (Xue *et al.*, 2026) (Turowski *et al.*, 2016). In particular, the strengthening of both initiation- and termination-associated elements may enhance transcriptional robustness while limiting interference with neighboring genes, especially those transcribed by Pol II (Turowski *et al.*, 2016). Beyond transcriptional efficiency, this coordinated reinforcement may contribute to stabilizing tDNA output in the context of genome expansion. Given the weak scaling between tDNA copy number and genome size in plants (Figure 4), modulation of *cis*-regulatory strength may serve as an important layer to ensure effective tRNA supply without requiring changes in gene copy number (Hummel *et al.*, 2019), (Monloy and Planta, 2024). Together, these observations support a model in which tDNA evolution in angiosperms rely on a dual regulatory strategy combining relatively stable gene dosage with optimized transcriptional activity through strengthened *cis*-regulatory elements.

At a more local scale, we observed reduced *cis*-regulatory divergence within clustered tDNAs compared to dispersed ones (Figure 6). However, the nature of this conservation varies among cluster types. In *A. thaliana*, the SYY cluster displays strong conservation of both upstream and downstream regions and preserves the internal organization of the repeated unit. This conservation occurs despite the heterochromatic context of this region (Hummel *et al.*, 2020) and the elevated mutation rates typically associated with closed chromatin (Makova and Hardison, 2015). Notably, the SYY cluster is selectively reactivated in root cap columella cells, where it has been proposed to supply $tRNA^{Ser}$ and $tRNA^{Tyr}$ for the translation of Ser/Tyr-rich proteins of the hydroxyproline-rich glycoprotein family known to be released by these cells in the mucilage (Hummel *et al.*, 2025). This specialized, cell-type-specific role may subject the SYY cluster to strong purifying selection, potentially counteracting the mutagenic effects of heterochromatin. The near *Brassicaceae*-specific distribution of this cluster further raises the possibility of a lineage-enriched functional specialization, potentially linked to the distinctive organization and secretory activity of the root cap in this family (Kumar and Iyer-Pascuzzi, 2020) (Kumpf and Nowack, 2015). Although this hypothesis remains highly speculative, such developmental constraints could contribute to the selective maintenance of the SYY cluster in this lineage. By contrast, the Pro cluster exhibits weaker conservation of upstream regions but more homogeneous downstream elements, pointing to stronger constraints on transcription termination. This pattern is consistent with its localization within intragenic euchromatic regions (Hummel *et al.*, 2020) associated with Pol II transcription, where chromatin accessibility and transcriptional interference may impose distinct regulatory constraints. Efficient termination in this context is likely critical to promote Pol III recycling and re-initiation while preventing read-through transcription and interference within tandem arrays (Yukawa *et al.*, 2000) (Dieci *et al.*, 2013) (Turowski and Tollervey, 2016). From a functional perspective, the conservation of Pro cluster architecture across dicot angiosperms (Figures 5 and 6) may reflect conditional translational demand for proline-rich proteins. Although largely silent under normal conditions in *A. thaliana* (Hummel *et al.*, 2020), this cluster may contribute to rapid synthesis of proline-rich proteins, such as extensins and hydroxyproline-rich glycoproteins (HRGPs), which are involved in cell wall reinforcement under stress conditions (Showalter *et al.*, 2010) (Johnson *et al.*, 2017) (Hummel *et al.*, 2025). More broadly, tDNA clustering appears to be the exception rather than the rule across photosynthetic eukaryotes and concerns only a limited subset of isotypes (Figure 5). The recurrence of similar cluster configurations across distantly related species further points to shared selective constraints that may favor the emergence and maintenance of these arrangements. Together, these observations suggest that tDNA clustering is neither random nor universal but instead reflects specific functional constraints and lineage-specific genomic dynamics.

Consistent with the limited extent of tDNA clustering, our analyses reveal that land plant genomes exhibit a predominantly dispersed organization of nuclear tDNAs, characterized by relatively regular inter-tDNA distances (Figure 4). Moreover, rather than scaling tDNA copy number with genome size, land plants primarily accommodate genomes expansion by increasing the spacing between tDNA loci. This pattern contrasts with that observed in many animals and fungi, where tDNA copy number often scales positively with genome size and inter-tDNA distances appear more heterogeneous (Goodenbour and Pan, 2006) (Bermudez-Santana *et al.*, 2010), (Santos and Del-Bem, 2023). Interestingly, a similarly irregular distribution of tDNA is also observed in the green alga *C. reinhardtii* (Figure S8), suggesting that the predominantly dispersed organization may represent a specific feature of land plant genomes rather than a general property of photosynthetic eukaryotes. The evolutionary forces shaping the distinct tDNA organization observed in land plants remain unclear but may partly reflect selective constraints linked to translational efficiency. Once an optimal tRNA supply matching codon demand is reached, further increases in tDNA copy number are likely disfavored, as they impose unnecessary energetic costs and increase the risk of translational imbalance (Goodenbour and Pan, 2006) (Du *et al.*, 2017). In sessile organisms such as land plants, where growth and stress responses depend on tight resource allocation under fluctuating environmental conditions, limiting unnecessary transcriptional and energetic costs may represent an additional selective advantage.

Taken together, our results suggest that nuclear tDNA organization in photosynthetic eukaryotes emerges from the interplay of multiple constraints operating across genomic scales. At the molecular level, selective pressures on translation shape tDNA repertoire composition and *cis*-regulatory elements controlling Pol III transcription. At the chromosomal level, genome architecture might govern spatial distribution, ensuring compatibility with chromatin environments and large-scale genome organization. These constraints are further modulated by lineage-specific evolutionary processes, including duplication, genome expansion, and ecological adaptation, which introduce additional variability in tDNA copy number and clustering patterns. This integrated framework reveals tDNAs as dynamic genomic elements positioned at the interface between translational demand, chromatin organization, and genome evolution. Rather than evolving independently, tDNA copy number, regulatory architecture, and spatial organization appear to be jointly optimized, defining a structured evolutionary landscape that underlies the diversification and robustness of translational systems across photosynthetic lineages.

## EXPERIMENTAL PROCEDURES

### Data sources

Nuclear tDNA annotations for photosynthetic eukaryotes were retrieved from the PlantRNA 2.0 database (http://plantrna.ibmp.cnrs.fr), (Cognat *et al.*, 2022), which provides curated tDNA annotations together with associated genomic features, including flanking sequences. The dataset comprises 49 representative species spanning the major Archaeplastida lineages (glaucophytes, rhodophytes, chlorophytes, streptophyte algae, bryophytes, lycophytes, gymnosperms, and angiosperms) as well as 4 lineages derived from secondary endosymbiosis (Table S1). To provide phylogenetic outgroups, tDNA annotations for five non-plant organisms were retrieved from the Genomic tRNA Database (http://gtrnadb.ucsc.edu), (Chan and Lowe, 2016) including *Saccharomyces cerevisiae* (S288c), *Drosophila melanogaster* (dm6), *Caenorhabditis elegans* (ce11), *Mus musculus* (mm39), *and Homo sapiens* (hg38). In addition, two non-plant eukaryotic lineages were included: the amoebozoan *Dictyostelium discoideum* and the excavate *Trypanosoma brucei*. For these species, nuclear tDNAs were predicted from genome assemblies (GCA_000004695.1 and GCF_000002445.2 respectively) using tRNAscan-SE v2.0 (Chan and Lowe, 2019) with default eukaryotic parameters. Annotation procedures followed those described in (Cognat et al., 2022). The final dataset comprised 60 species and a total of 38,233 nuclear tDNAs.

### Computational meta-analysis of tDNA organization

Custom R scripts were developed to process and integrate tDNA annotations from PlantRNA datasets across species. Only nuclear-encoded tDNAs were retained. All entries corresponding to mitochondrial or plastidial tDNAs were excluded. Scaffold sequences, or unplaced contigs were excluded from genome organization analysis. All analyses were performed in R (v4.5.0). Data processing and genomic interval analyses relied on the packages *data.table* (v1.17.8), *GenomicRanges*, and *rtracklayer*, and graphical outputs were generated using *ggplot2* (v3.5.2), *ggseqlogo* and *ggtree*. For each species, genomic coordinates of nuclear tDNAs were converted into strand-aware GRanges objects and ordered along chromosomes. Inter-tDNA distance (*d*) was defined as the genomic distance between consecutive tDNAs along each chromosome, computed using rank-ordered coordinates. Distances were calculated genome-wide and per chromosome, and log10-transformed for visualization. Chromosome-level distributions of tDNA isotypes were assessed by comparing observed counts per chromosome to expected counts based on chromosome length. Expected values were calculated assuming a uniform distribution proportional to chromosome size. Deviations from expectation were quantified using standardized residuals derived from chi-square statistics. Heatmap values correspond to standardized Pearson residuals comparing observed and expected chromosome-level isotype counts. Deviations were considered significant when exceeding the chi-square threshold for 1 degree of freedom ($\chi^2 > 3.84$; corresponding to |standardized Pearson residual| > 1.96), allowing classification of chromosome-specific enrichment or depletion. Pairwise adjacency relationships between consecutive tDNAs were also computed genome-wide to characterize local neighborhood organization. Detailed statistical procedures for adjacency analyses are provided in the Supplementary Methods. To characterize local genomic organization, tDNA clusters were defined using proximity-based criteria. Unless otherwise stated, clusters were defined as groups of at least five tDNAs separated by ≤5 kb, a threshold selected as a robust compromise between sensitivity and specificity for detecting compact clusters across species. Alternative thresholds (≥3 genes within 1 kb; ≥10 genes within 50 kb) were also evaluated to assess the robustness of clustering patterns (Figure S12). For each species, clustering ratios were calculated as the proportion of tDNAs assigned to clusters relative to the total number of annotated nuclear tDNAs. Detailed formulas and implementation procedures are provided in Supplementary Methods.

**Extraction of flanking sequences and *cis*-regulatory feature analyses**

To investigate *cis*-regulatory features associated with Pol III transcription, genomic flanking regions were extracted for each annotated nuclear tDNA using strand-aware coordinates. For each gene, 50 nucleotides upstream (-50 to -1 relative to the 5′ end of the tDNA coding sequence) and 25 nucleotides downstream (+1 to +25 relative to the 3′ end) were retrieved from the corresponding genome assembly. Upstream and downstream sequences were analyzed separately to quantify features associated with transcriptional initiation and termination. Sequence logos were generated from aligned flanking regions using the *ggseqlogo* package. Three *cis*-regulatory features were quantified for each locus: (*i*) upstream AT enrichment, calculated as the proportion of adenine and thymine nucleotides within the -35 to -20 region; (*ii*) CAA motif occurrence and positional bias, assessed by recording both the presence or absence of at least one CAA trinucleotide within the −10 to −3 upstream region and, when present, the position of the first detected motif; and (*iii*) downstream termination features, including total thymidine content within the +1 to +25 region and the length of the longest contiguous poly(T) stretch. For comparative analyses, species were grouped into five major phylogenetic categories: angiosperms, non-angiosperm embryophytes, other Archaeplastida, photosynthetic eukaryotes with secondary plastids, and non-photosynthetic eukaryotes (Figure 1a, Table S1). Group-wise comparisons of *cis*-regulatory features were performed using Kruskal-Wallis tests followed by pairwise Wilcoxon rank-sum tests with Benjamini-Hochberg correction for multiple testing. Complete statistical results are provided in Table S2. To assess whether *cis*-regulatory patterns differed according to genomic context, clustered and dispersed tDNAs were analyzed separately for representative isotype families. Flanking sequence similarity was quantified using uncorrected p-distances calculated with the *dist.dna* function from the *ape* package (model = "N"). Distance trees were inferred using the BIONJ algorithm and visualized with *ggtree*.

**ShinytRNA web application**

ShinytRNA was developed as an interactive web application in R using the shiny framework (v1.11.1) to facilitate standardized exploration of nuclear tDNA genomic organization across species. The application was implemented using *data.table* (v1.17.8), *ggplot2* (v3.5.2), and *DT* (v0.33) for interactive visualization and export of tabular outputs. ShinytRNA integrates five complementary modules enabling genome-scale analysis of tDNA organization: (*i*) a tDNA isotype distribution module for chromosome-level enrichment analyses, (*ii*) an inter-tDNA distance module for spatial dispersion analyses, (*iii*) an adjacency (pairwise) module for local neighborhood analyses, (*iv*) a tDNA cluster module for proximity-based cluster detection, and (*v*) a centromere positioning module, based on experimentally validated centromere coordinates when available or proxy inference from large tDNA-free genomic intervals otherwise (Figure 1b; Figure S1). The application allows dynamic parameter adjustment and export of publication-ready graphical and tabulated outputs.

## ACKNOWLEDGMENTS

This work was supported by the Centre National de la Recherche Scientifique (CNRS).

## AUTHOR CONTRIBUTIONS

GH designed work, performed bioinformatic analyses, analyzed the data, and participated in writing the manuscript. DP developed the ShinytRNA platform, performed bioinformatics analyses and participated in writing the manuscript. VC participated in tDNA annotation and processing of genomes. LD and AB supervised the work, designed experiments, analyzed the data and wrote the manuscript.

## CONFLICT OF INTEREST STATEMENT

The authors reported no potential conflict of interest.

## DATA AVAILABILITY STATEMENT

ShinytRNA is freely accessible at: https://nebula.ibmp.unistra.fr/shinytRNA/.

All scripts and processed datasets are available at [https://github.com/ibmp/tDNA-organization-meta-analysis] to ensure full reproducibility.

All the relevant data can be found within the manuscript and its supporting materials.

## SUPPORTING INFORMATION

Additional Supporting information may be found in the online version of this article.

**Supplemental methods**

**Figure S1.** Data analysis performed using the ShinytRNA platform https://nebula.ibmp.unistra.fr/shinytRNA/).

**Figure S2**. Boxplot showing the CAA start position of the first CAA motif in the region upstream of the 5′ end of tDNAs.

**Figure S3.** Histograms showing the proportion of tDNAs with a CAA motif within the 10 to -3 upstream region

**Figure S4.** Boxplot showing the AT content in the region from –50 to –35 bp upstream of the 5′ end of tDNAs.

**Figure S5.** Boxplot showing the T content within 25 bp downstream of the 3′ end of tDNAs.

**Figure S6.** Boxplot showing the length of the longest T stretch located within 25 bp downstream of the 3′ end of tDNAs.

**Figure S7.** Boxplots showing the distance (*d*) between consecutive tDNAs on chromosomes from land plant species.

**Figure S8:** Comparison of tDNA organization in *Chlamydomonas reinhardtii* versus a representative plant species (*Nicotiana tabacum*) and two non-plant species (*Mus musculus* and *Homo sapiens*).

**Figure S9**. tDNA organization in the green lineage including *Chlamydomonas reinhardtii*.

**Figure S10.** Heatmaps showing the chromosomal distribution of tDNA isotypes in three species.

**Figure S11.** Chromosomal distribution of tDNAs reveals large-scale genomic organization in plants

**Figure S12.** tDNA cluster identification upon different filtering conditions.

**Figure S13.** Identification of proline-, serine-, and tyrosine-containing tDNA clusters.

**Table S1.** Photosynthetic and non-photosynthetic species analyzed in this study.

**Table S2.** All Pairwise Wilcoxon Rank Sum Tests with Benjamini & Hochberg (BH) adjustment.

**Table S3.** Detailed annotation of tRNA gene clusters identified under different filtering conditions.

## LEGEND TO FIGURES

**Figure 1: Datasets and analytical frameworks used in this study.** a) Schematic phylogenetic representation of the species included in this study. For photosynthetic organisms, the tree derives from (Cognat et al., 2022). Seven non-photosynthetic eukaryotes were also included for comparison. For several analyses, species were grouped into five major categories: angiosperms, non-angiosperm embryophytes, other Archaeplastida, photosynthetic eukaryotes with secondary plastids, and non-photosynthetic eukaryotes. LECA: Last Eukaryotic Common Ancestor. Three-letter species abbreviations are defined in Table S1. b) Overview of the ShinytRNA platform (https://nebula.ibmp.unistra.fr/shinytRNA/) developed for comparative analysis of nuclear tDNA organization. ShinytRNA integrates complementary modules enabling genome-wide exploration of tDNA organization. The tDNA Isotype module quantifies amino acid-specific distributions and tests enrichment across chromosomes. tDNA Distribution evaluates spatial organization by measuring distances

between consecutive tDNA. tDNA Pairwise identifies over- and under-represented transitions between neighboring tDNA. tDNA Clusters detects genomic clusters using adjustable thresholds and reports their composition and density. Centromere Positioning infers putative centromere locations based on large inter-tDNA gaps and visualizes their distribution along chromosomes.

**Figure 2: Conserved distribution of tDNA copy-numbers across eukaryotes.** Box plot (top) and heatmap (bottom) showing the distribution of nuclear tDNA copy numbers per amino acid across representative photosynthetic organisms and other eukaryotes. Amino acids are ordered according to amino acid frequency in *A. thaliana* (Michaud et al., 2011). Elongator methionine (eMet) and initiator methionine (iMet) tRNAs are distinguished. Clustered tDNAs were excluded. Names of photosynthetic organisms are abbreviated as in Table S1.

**Figure 3: Comparative analysis of *cis*-regulatory elements associated with tDNA transcription.** a) Schematic representation of the major *cis*-regulatory elements associated with RNA polymerase III (Pol III) transcription of tDNAs. Upstream regions include A/T-rich sequences and CAA motifs, while internal A and B boxes and downstream poly-T stretches define the core Pol III promoter architecture. These elements interact with transcription factors TFIIIC and TFIIIB to ensure transcriptional initiation, elongation, and termination. b) Sequence logos showing nucleotide conservation in upstream (−50 to −1) and downstream (+1 to +25) flanking regions of tDNAs from *Arabidopsis thaliana* (Ath), *Marchantia polymorpha* (Mpo), *Physcomitrium patens* (Ppa), and *Chlamydomonas reinhardtii* (Cre). The +1 position marks the first nucleotide of the mature tRNA. c) Boxplots comparing *cis*-regulatory sequence features across five major organismal groups: (A) angiosperms (n = 16,367 tDNAs), (B) non-angiosperm embryophytes (n = 8,091), (C) other Archaeplastida (n = 2,018), (D) photosynthetic eukaryotes with secondary plastids (n = 411), and (E) non-photosynthetic eukaryotes (n = 1,298) (see Figure S2 for details). Parameters include the position of the first upstream CAA motif, upstream A/T content, the length of the longest downstream poly-T stretch, and downstream T enrichment. Statistical significance relative to angiosperms is indicated by horizontal bars (*** $P < 0.001$; see Supplementary Table S2 and Figures S8-S11).

**Figure 4: Chromosomal dispersion and spacing of tDNAs across plant genomes.** a) Boxplots showing intergenic distances (*d*) between adjacent tDNAs (trn-1 to trn) along *A. thaliana* chromosomes, calculated either including (left) or excluding (right) tDNA clusters. b) Variation in intergenic distances between adjacent tDNAs across species. Boxplots compare (i) all chromosomes (blue), (ii) all chromosomes excluding clusters (orange), (iii) only chromosomes containing at least one tDNA cluster, with intra-cluster distances included (red),

and (iv) the same set of chromosomes after excluding intra-cluster distances (purple). Species are ordered by increasing genome size and species lacking clusters are omitted. c) Linear regressions (dark grey lines) depicting the relationship between genome size and total tDNA number (left) or median intergenic distance between tDNAs (right). Regression equations and coefficients of determination ($R^2$) are indicated. Grey shaded areas represent the 95% confidence interval. Each point corresponds to one species. *Chlamydomonas reinhardtii* is excluded here but shown in Figure S7. d) Heatmaps illustrating the chromosomal distribution of isotype families in *Marchantia polymorpha* (Mpo), *Arabidopsis thaliana* (Ath), and *Nicotiana tabacum* (Nta). Rows represent chromosomes and columns amino acid families (single-letter code). Colors indicate deviation from a random distribution normalized to genome size, with red indicating enrichment and blue depletion. Clustered tDNAs were excluded.

**Figure 5: Comparative analysis of tDNA clustering across species and isotype families.** Histograms showing the percentage of total clusterized tDNAs in each species (left) and heatmap showing the clustering ratio of tDNA isotypes in each species (right), under the clustering criterion ≥5 genes within 5 kb. White indicates absence of clustering (all genes isolated), whereas dark red indicates complete clustering (all genes grouped), with the color scale representing clustering ratio (%) from 0 to 100. White arrows point to three tDNA isotype families (proline/P, serine/S, and tyrosine/Y) that frequently cluster in angiosperms (see blue and red histograms). In *C. merolae* (Cme) and *P. tricornutum* (Phatr), asparagine (D) and cysteine (C) tDNAs could not be confidently annotated and were excluded from the analysis (black crossed-out boxes). Complete results for all clustering criteria are provided in Figure S11.

**Figure 6: Clustered tDNAs exhibit increased *cis*-regulatory sequence conservation relative to dispersed loci.** a) Phylogenetic trees generated from flanking sequences of nuclear tDNAs together with the corresponding sequence logos illustrating nucleotide conservation within these regions. For each example, loci belonging to clusters are shown in blue, whereas dispersed loci are shown in black. Trees were constructed separately from upstream (Up) and downstream (Down) flanking sequences, and the associated sequence logos summarize nucleotide frequencies within the same regions. Upstream sequences correspond to positions −50 to −1 relative to the tDNA start, whereas downstream sequences correspond to +1 to +25 relative to the 3′ end of the tDNA coding sequence. The representative clusters include $\text{tDNAs}^{\text{Pro}}$ in *Arabidopsis thaliana* (Ath), $\text{tDNAs}^{\text{Ile}}$ in *Zea mays* (Zma), and $\text{tDNAs}^{\text{Gln}}$ in *Hordeum vulgare*, which are organized as tandem arrays of the same isotype, as well as the Ser-Tyr-Tyr cluster in *A. thaliana*, corresponding to tandem arrays of a repeated Ser-Tyr-Tyr module. Within this module, the first Tyr copy groups toward the top of the phylogenetic tree, whereas the second Tyr copy groups toward the bottom, reflecting their distinct positions within the repeated motif.

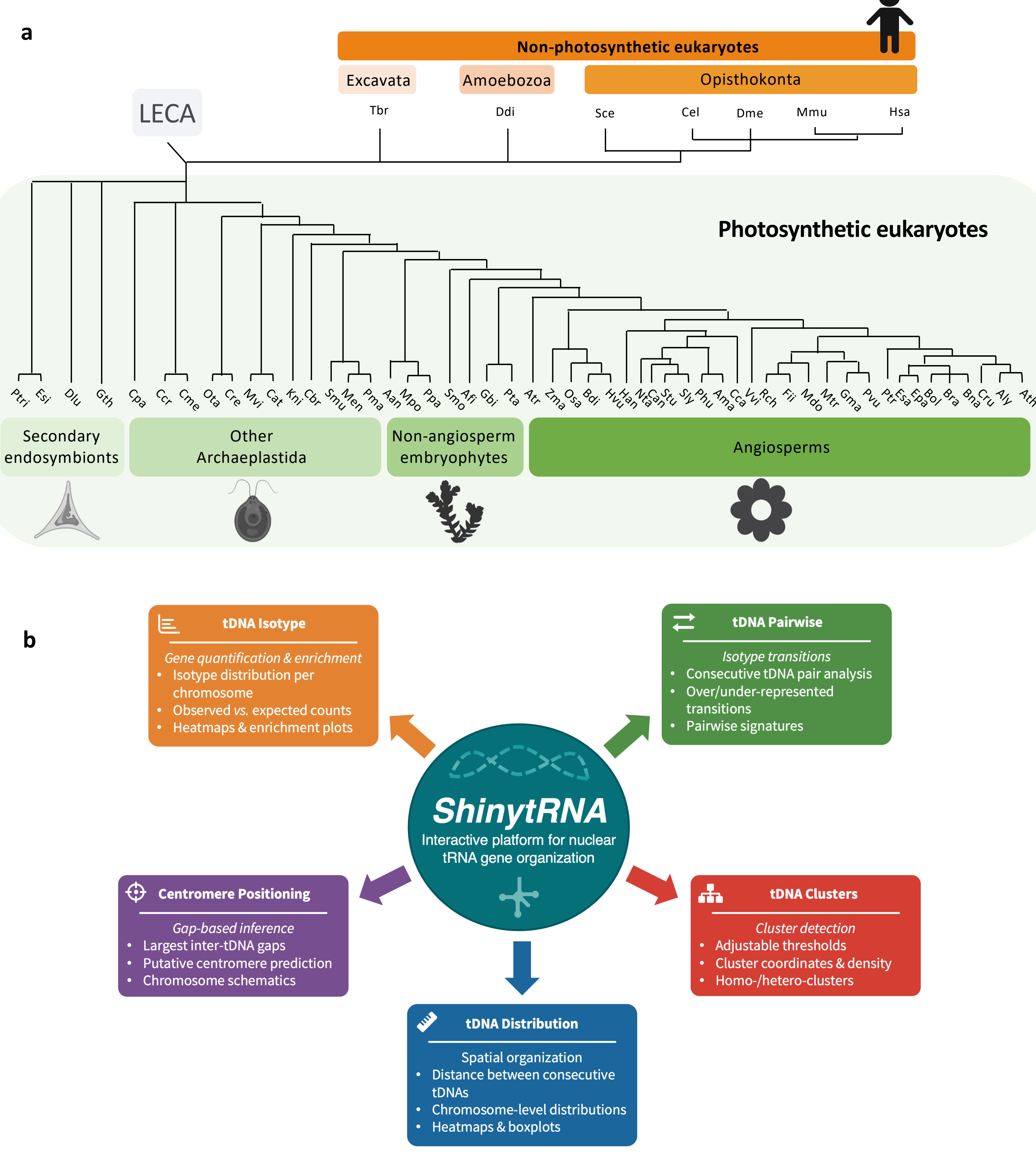

**Figure 1: Datasets and analytical frameworks used in this study. a)** Schematic phylogenetic representation of the species included in this study. For photosynthetic organisms, the tree derives from (Cognat et al., 2022). Seven non-photosynthetic eukaryotes were also included for comparison. For several analyses, species were grouped into five major categories: angiosperms, non-angiosperm embryophytes, other Archaeplastida, photosynthetic eukaryotes with secondary plastids, and non-photosynthetic eukaryotes. LECA: Last Eukaryotic Common Ancestor. Three-letter species abbreviations are defined in Table S1. **b**) Overview of the ShinytRNA platform (https://nebula.ibmp.unistra.fr/shinytRNA/) developed for comparative analysis of nuclear tDNA organization. ShinytRNA integrates complementary modules enabling genome-wide exploration of tDNA organization. The tDNA Isotype module quantifies amino acid-specific distributions and tests enrichment across chromosomes. tDNA Distribution evaluates spatial organization by measuring distances between consecutive tDNA. tDNA Pairwise identifies over- and under-represented transitions between neighboring tDNA. tDNA Clusters detects genomic clusters using adjustable thresholds and reports their composition and density. Centromere Positioning infers putative centromere locations based on large inter-tDNA gaps and visualizes their distribution along chromosomes.

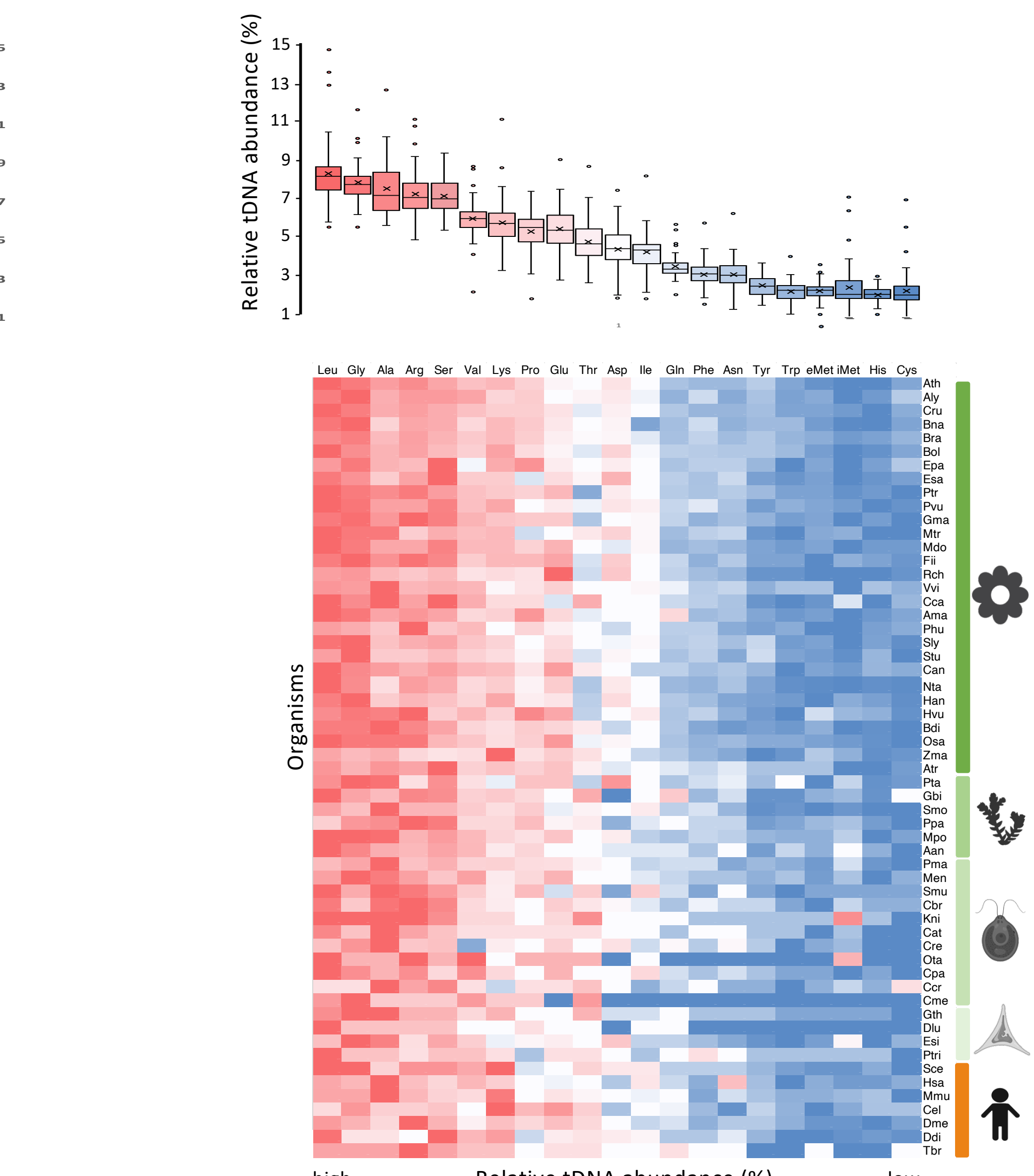

**Figure 2: Conserved distribution of tDNA copy-numbers across eukaryotes.** Box plot (top) and heatmap (bottom) showing the distribution of nuclear tDNA copy numbers per amino acid across representative photosynthetic organisms and other eukaryotes. Amino acids are ordered according to amino acid frequency in *A. thaliana* -(Michaud et al., 2011). Elongator methionine (eMet) and initiator methionine (iMet) tRNAs are distinguished. Clustered tDNAs were excluded. Names of photosynthetic organisms are abbreviated as in Table S1.

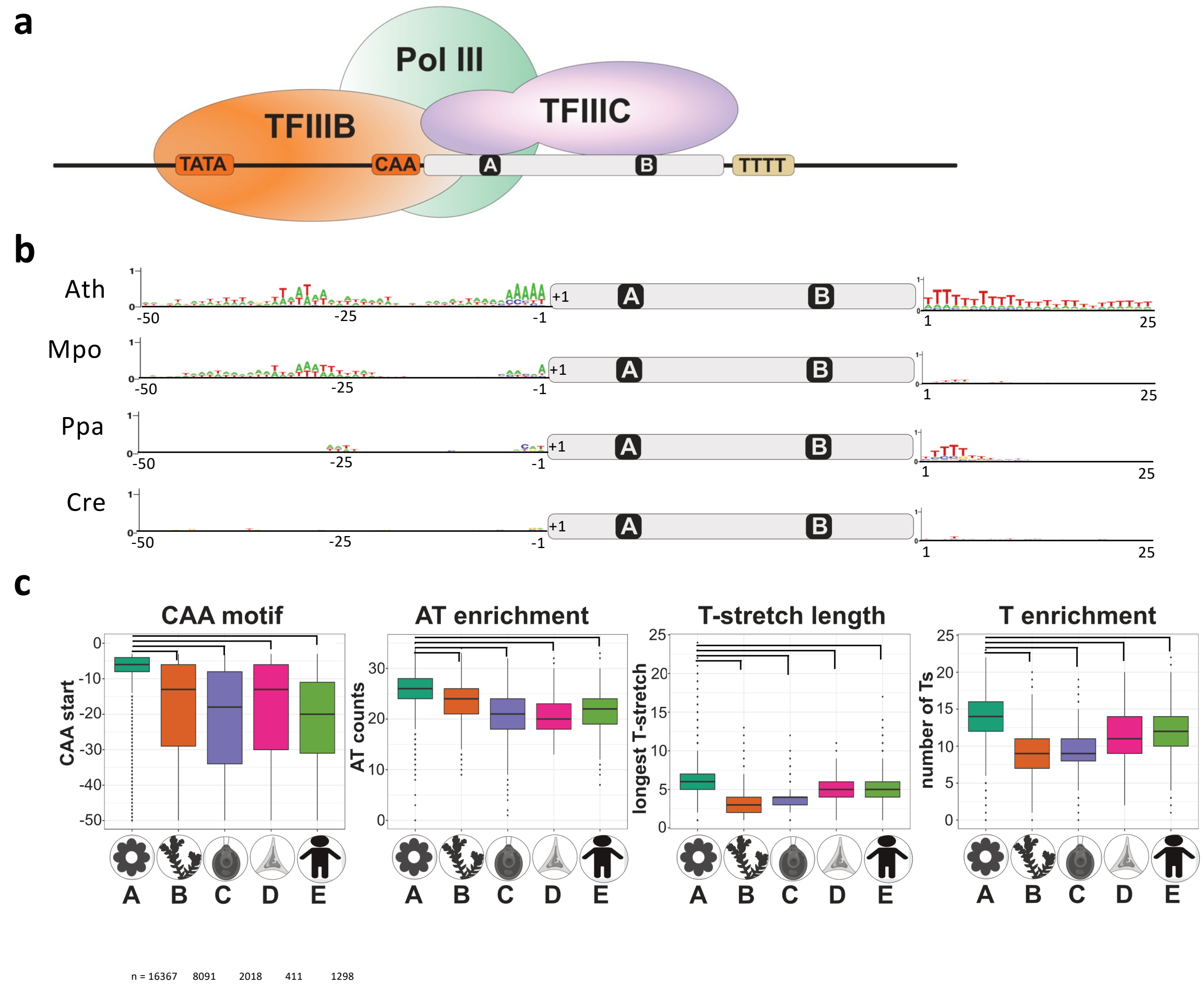

**Figure 3: Comparative analysis of *cis*-regulatory elements associated with tDNA transcription. a**) Schematic representation of the major *cis*-regulatory elements associated with RNA polymerase III (Pol III) transcription of tDNAs. Upstream regions include A/T-rich sequences and CAA motifs, while internal A and B boxes and downstream poly-T stretches define the core Pol III promoter architecture. These elements interact with transcription factors TFIIIC and TFIIIB to ensure transcriptional initiation, elongation, and termination. **b**) Sequence logos showing nucleotide conservation in upstream (−50 to −1 ) and downstream (+1 to +25) flanking regions of tDNAs from *Arabidopsis thaliana* (Ath), *Marchantia polymorpha* (Mpo), *Physcomitrium patens* (Ppa), and *Chlamydomonas reinhardtii* (Cre). The +1 position marks the first nucleotide of the mature tRNA. **c**) Boxplots comparing *cis*-regulatory sequence features across five major organismal groups: (A) angiosperms (n = 16,367 tDNAs), (B) non-angiosperm embryophytes (n = 8,091), (C) other Archaeplastida (n = 2,018), (D) photosynthetic eukaryotes with secondary plastids (n = 411), and (E) non-photosynthetic eukaryotes (n = 1,298) (see Figure S2 for details). Parameters include the position of the first upstream CAA motif, upstream A/T content, the length of the longest downstream poly-T stretch, and downstream T enrichment. Statistical significance relative to angiosperms is indicated by horizontal bars (*** $P < 0.001$; see Supplementary Table S2 and Figures S8-S11).

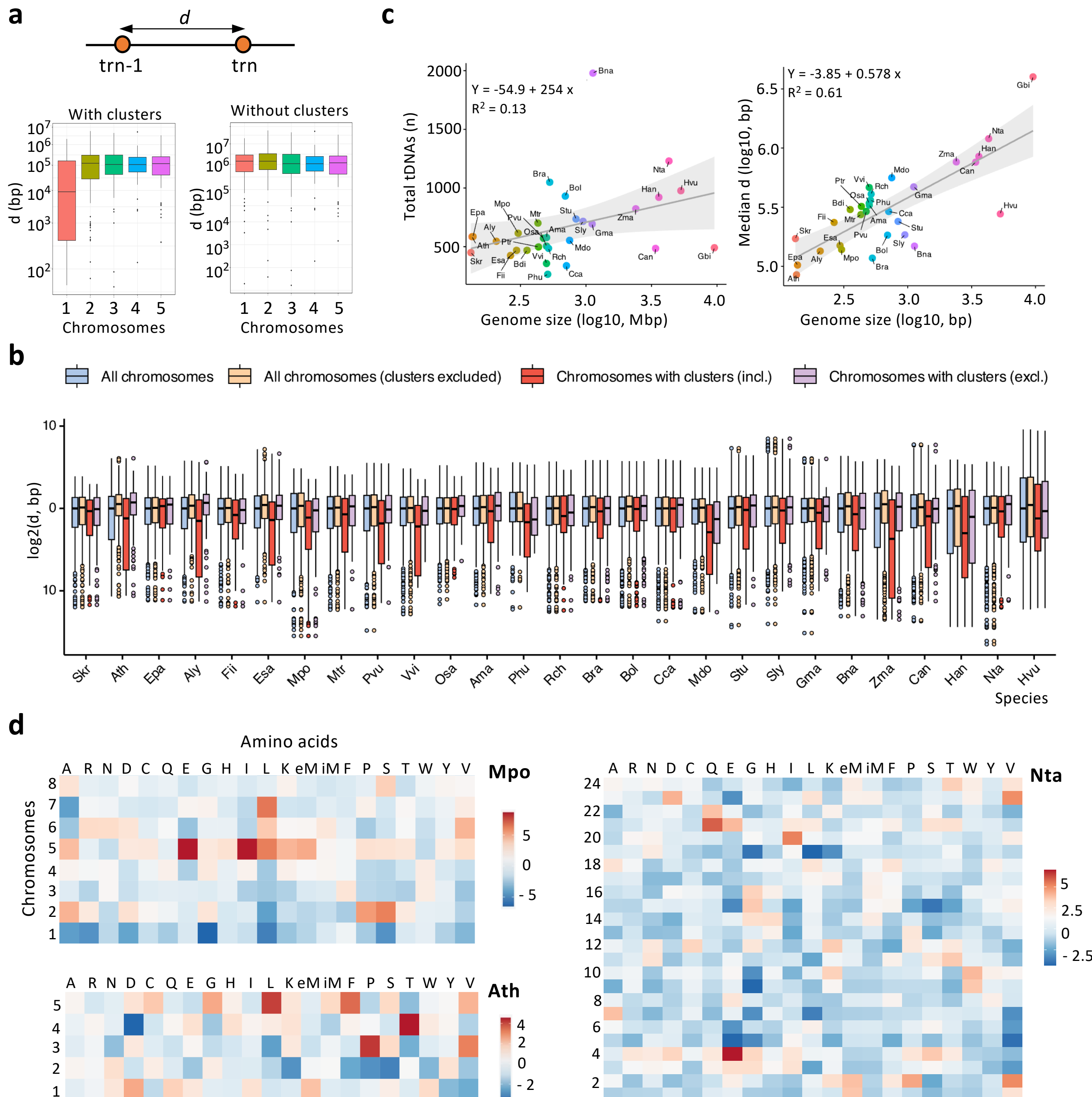

**Figure 4: Chromosomal dispersion and spacing of tDNAs across plant genomes. a**) Boxplots showing intergenic distances (*d*) between adjacent tDNAs (trn-1 to trn) along *A. thaliana* chromosomes, calculated either including (left) or excluding (right) tDNA clusters. **b**) Variation in intergenic distances between adjacent tDNAs across species. Boxplots compare (i) all chromosomes (blue), (ii) all chromosomes excluding clusters (orange), (iii) only chromosomes containing at least one tDNA cluster, with intra-cluster distances included (red), and (iv) the same set of chromosomes after excluding intra-cluster distances (purple). Species are ordered by increasing genome size and species lacking clusters are omitted. **c**) Linear regressions (dark grey lines) depicting the relationship between genome size and total tDNA number (left) or median intergenic distance between tDNAs (right). Regression equations and coefficients of determination ($R^2$) are indicated. Grey shaded areas represent the 95% confidence interval. Each point corresponds to one species. *Chlamydomonas reinhardtii* is excluded here but shown in Supplementary Figure S7. **d**) Heatmaps illustrating the chromosomal distribution of isotype families in *Marchantia polymorpha* (Mpo), *Arabidopsis thaliana* (Ath), and *Nicotiana tabacum* (Nta). Rows represent chromosomes and columns amino acid families (single-letter code). Colors indicate deviation from a random distribution normalized to genome size, with red indicating enrichment and blue depletion. Clustered tDNAs were excluded.

Photosynthetic pro
Non-photosynthetic
Fungi
Metazoans
Bryophytes
Lycophytes
Gymnosperms
Angiosperms

(%)
30

P S Y

A C D E F G H I K L eM iM N P Q R S T V W Y

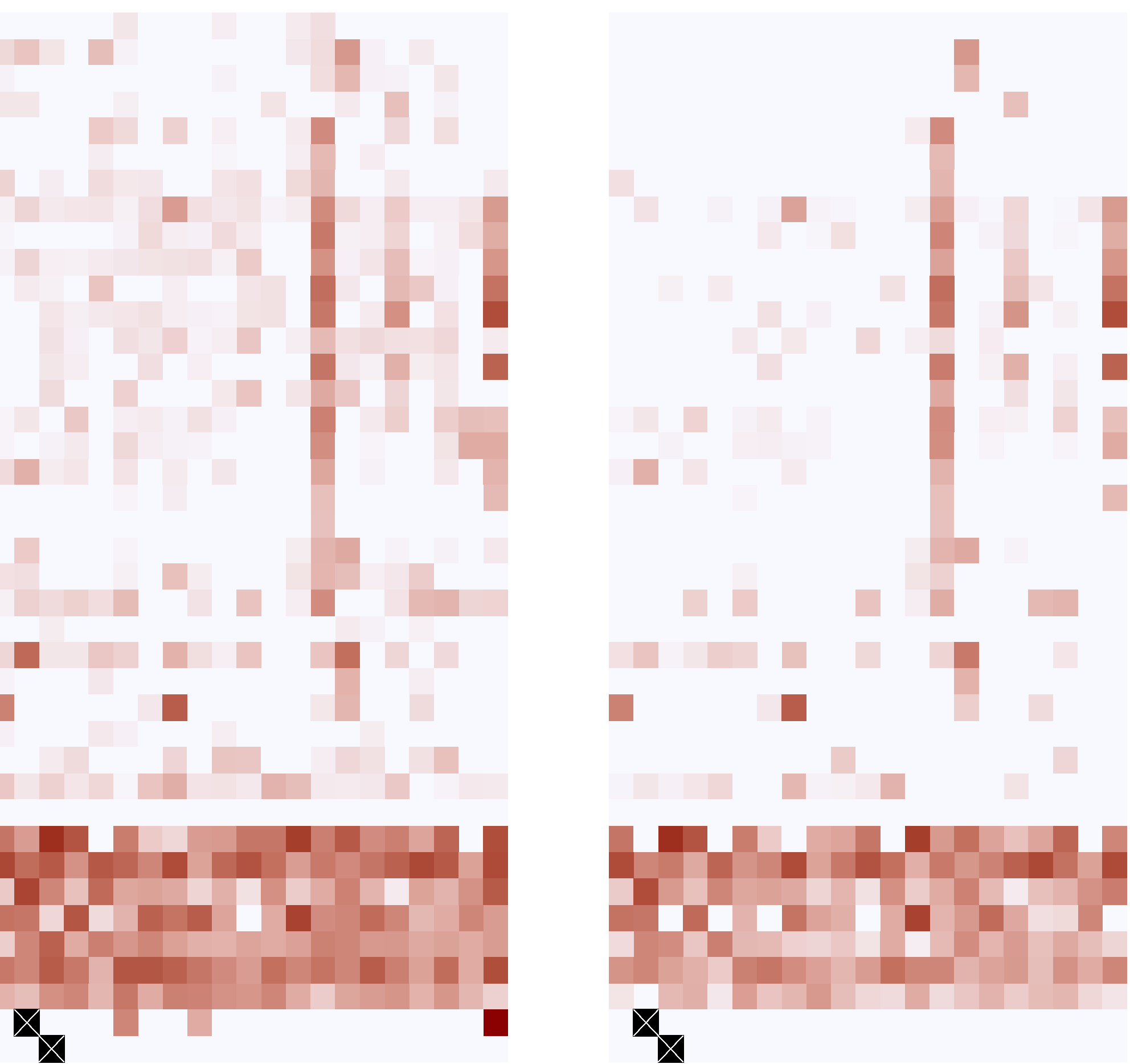

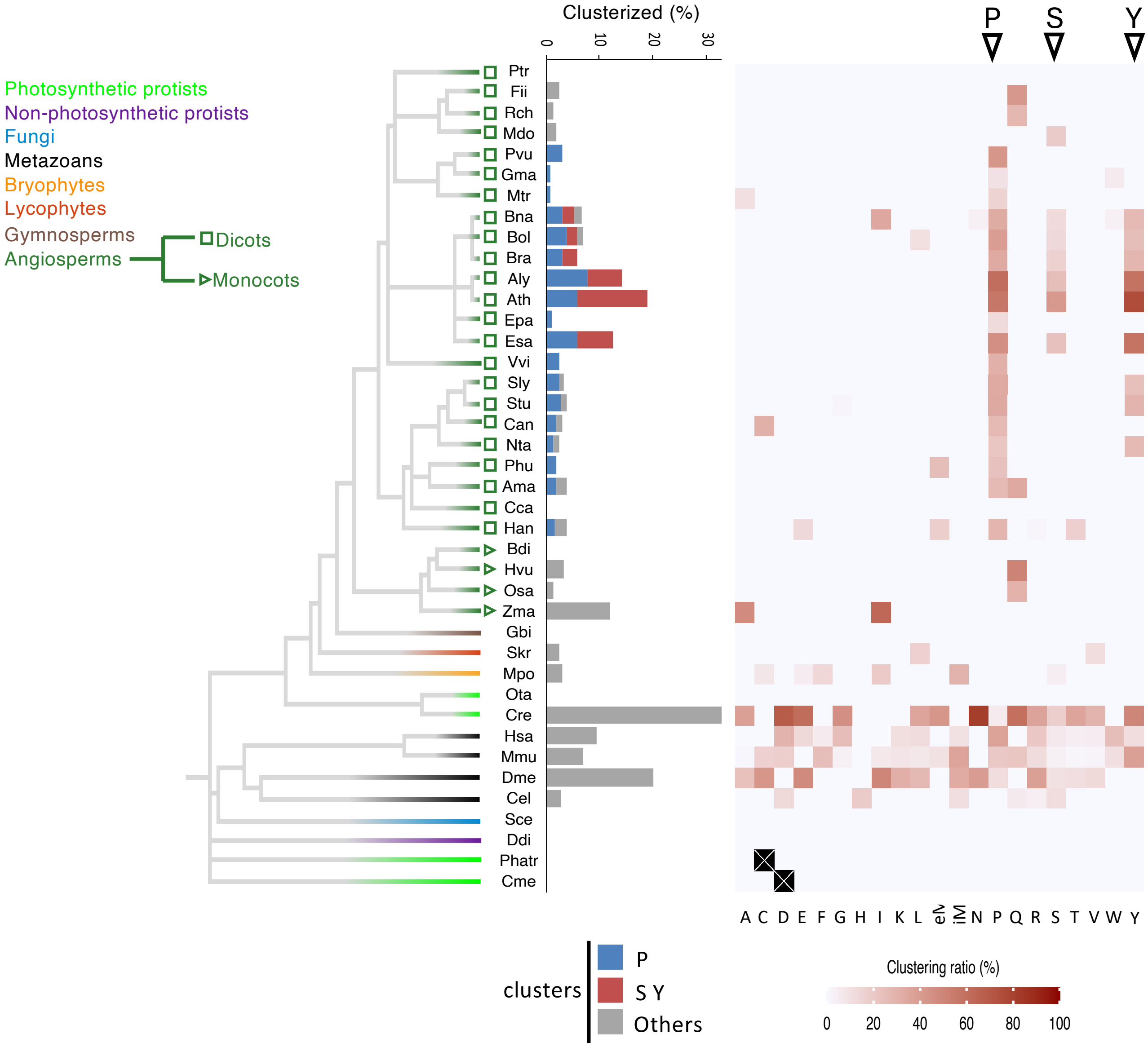

**Figure 5: Comparativ**
the percentage of to
tDNA isotypes in eac
absence of clustering
with the color scale r
families (proline/P, s
histograms). In *C. me*
be confidently annot
for all clustering crite

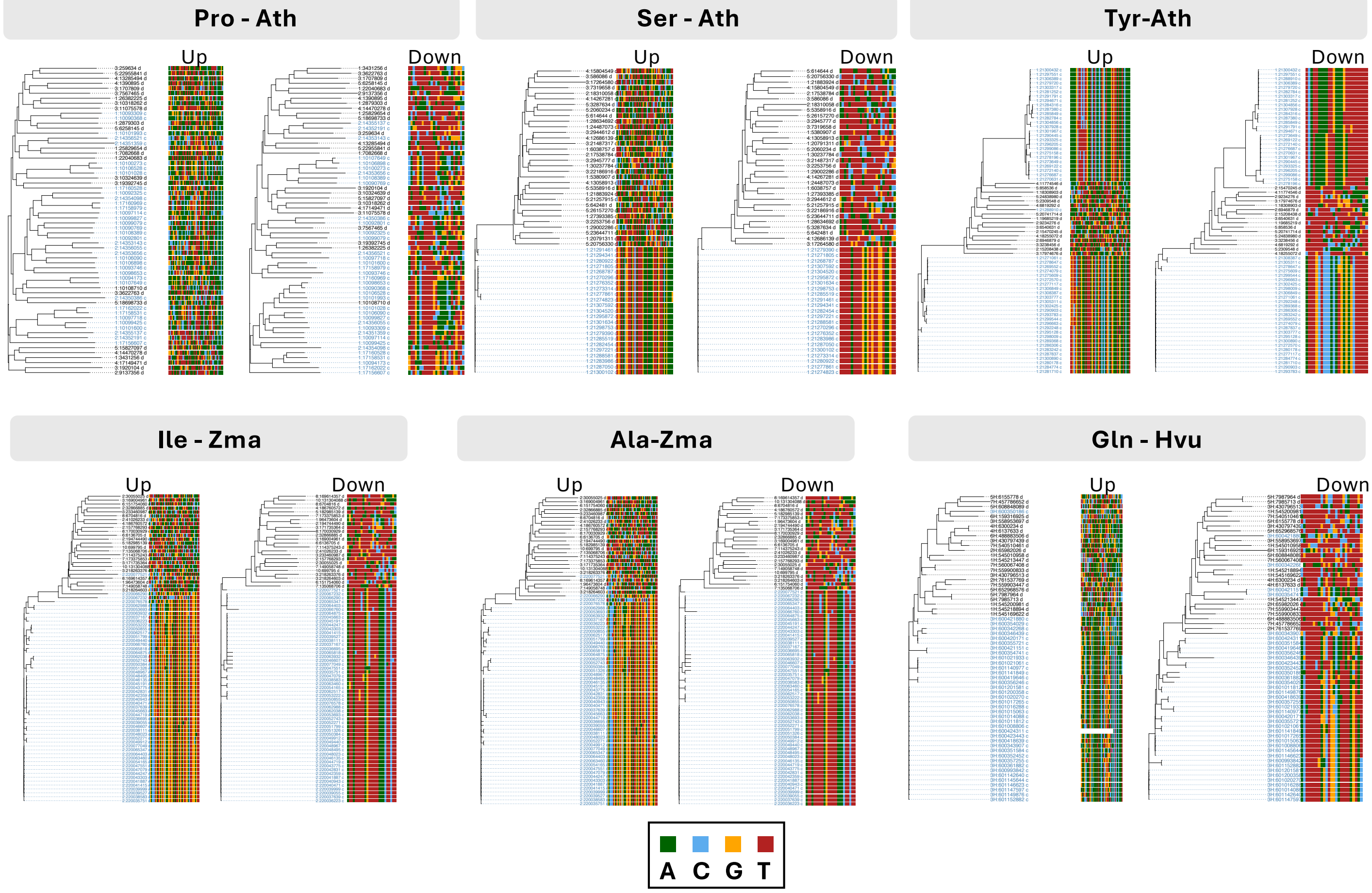

**Figure 6: Clustered tDNAs exhibit increased *cis*-regulatory sequence conservation relative to dispersed loci. a)** Phylogenetic trees generated from flanking sequences of nuclear tDNAs together with the corresponding sequence logos illustrating nucleotide conservation within these regions. For each example, loci belonging to clusters are shown in blue, whereas dispersed loci are shown in black. Trees were constructed separately from upstream (Up) and downstream (Down) flanking sequences, and the associated sequence logos summarize nucleotide frequencies within the same regions. Upstream sequences correspond to positions −50 to −1 relative to the tDNA start, whereas downstream sequences correspond to +1 to +25 relative to the 3′ end of the tDNA coding sequence. The representative clusters include tDNAs$^{Pro}$ in *Arabidopsis thaliana* (Ath), tDNAs$^{Ile}$ in *Zea mays* (Zma), and tDNAs$^{Gln}$ in *Hordeum vulgare*, which are organized as tandem arrays of the same isotype, as well as the Ser-Tyr-Tyr cluster in *A. thaliana*, corresponding to tandem arrays of a repeated Ser-Tyr-Tyr module. Within this module, the first Tyr copy groups toward the top of the phylogenetic tree, whereas the second Tyr copy groups toward the bottom, reflecting their distinct positions within the repeated motif.

# Supplemental Methods

## Detailed computational procedures

The analytical framework implemented in this study was designed to quantify nuclear tDNA organization across genomic scales, including intergenic spacing, chromosome-level distribution, adjacency relationships, local clustering, and *cis*-regulatory sequence conservation. Detailed formulas, threshold definitions, and algorithmic procedures used for each analysis are provided below.

## Inter-tDNA distance calculation

Inter-tDNA distance (*d*) was defined as the genomic distance between two consecutive tDNAs ($trn_i$ and $trn_{i+1}$) along a chromosome:

$$d = start\,(trn_{i+1}) - end\,(trn_i)$$

Distances were computed for all adjacent pairs within each chromosome. When indicated, loci belonging to clusters were excluded prior to distance calculation to estimate background dispersion. Distance distributions were summarized using median, mean, standard deviation, and extrema. For visualization, values were log10-transformed.

## Chromosome-level expected distribution model

To assess whether tDNA isotypes are randomly distributed across chromosomes, expected counts were calculated assuming proportionality to chromosome length:

$$E_{i,a} = \frac{N_a}{L_{genome}} \times L_i$$

where: $E_{i,a}$ = expected number of tDNAs of isotype *a* on chromosome *i* ; $N_a$ = expected number of tDNAs of isotype *a* on chromosome *I* ; $L_i$ = length of chromosome *I* ; and $L_{genome}$ = total genome size

Deviation from expectation was quantified using standardized residuals:

$$R_{i,a} = \frac{O_{i,a} - E_{i,a}}{\sqrt{E_{i,a}}}$$

where $O_{i,a}$ is the observed count.

Significance of deviations from expected distributions was assessed using chi-square tests with 1 degree of freedom ($p < 0.05$; $\chi^2 > 3.84$). Isotypes were classified as over-represented, under-represented, or neutral based on fold-change and statistical significance. In addition, a global non-randomness test was performed per isotype by aggregating chromosome-level chi-square statistics.

**Adjacency and pairwise isotype analysis**

To evaluate non-random organization of neighbouring tDNAs, pairwise adjacency frequencies were computed based on ordered loci along chromosomes.

Expected frequencies under a random model were calculated as:

For identical isotypes (a = b):

$$E_{a,a} = \frac{N_a(N_a - 1)}{N(N - 1)} \times T$$

For distinct isotypes (a ≠ b):

$$E_{a,b} = \frac{2N_a N_b}{N(N - 1)} \times T$$

where: $N_a$ and $N_b$ = counts of isotype *a* and *b* ; $N$ = total number of tDNAs ; $T$ = total number of adjacent pairs.

For each pair, the following metrics were computed:

- observed and expected counts
- observed and expected frequencies (%)
- fold enrichment
- frequency deviation (Δ)
- chi-square statistic and associated p-value

Observed and expected values were compared using chi-square tests with 1 degree of freedom.

**Centromere position approximation**

Putative centromeric regions were approximated based on large inter-tDNA distances. For each chromosome, the largest intergenic intervals were identified, and midpoint positions between adjacent loci were calculated as:

$$C = end_{n-1} + \frac{d}{2}$$

These regions were interpreted as low-density genomic intervals potentially corresponding to centromeric regions. When experimentally validated centromere coordinates were unavailable, putative centromeric regions were approximated from the largest inter-tDNA intervals. These proxy regions were used to assess large-scale tDNA exclusion patterns and should be interpreted cautiously.

## tRNA Isotype

Isotype distribution per chromosome

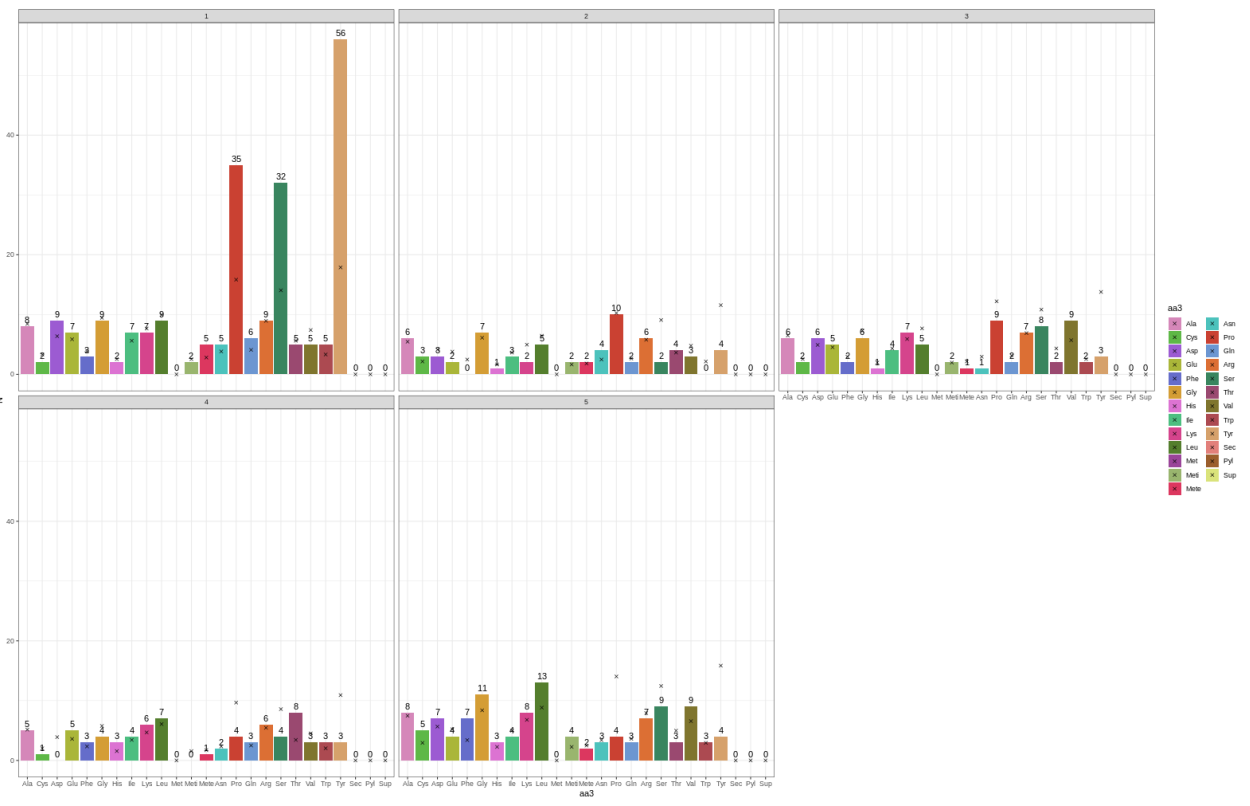

Heatmap of observed vs. expected counts

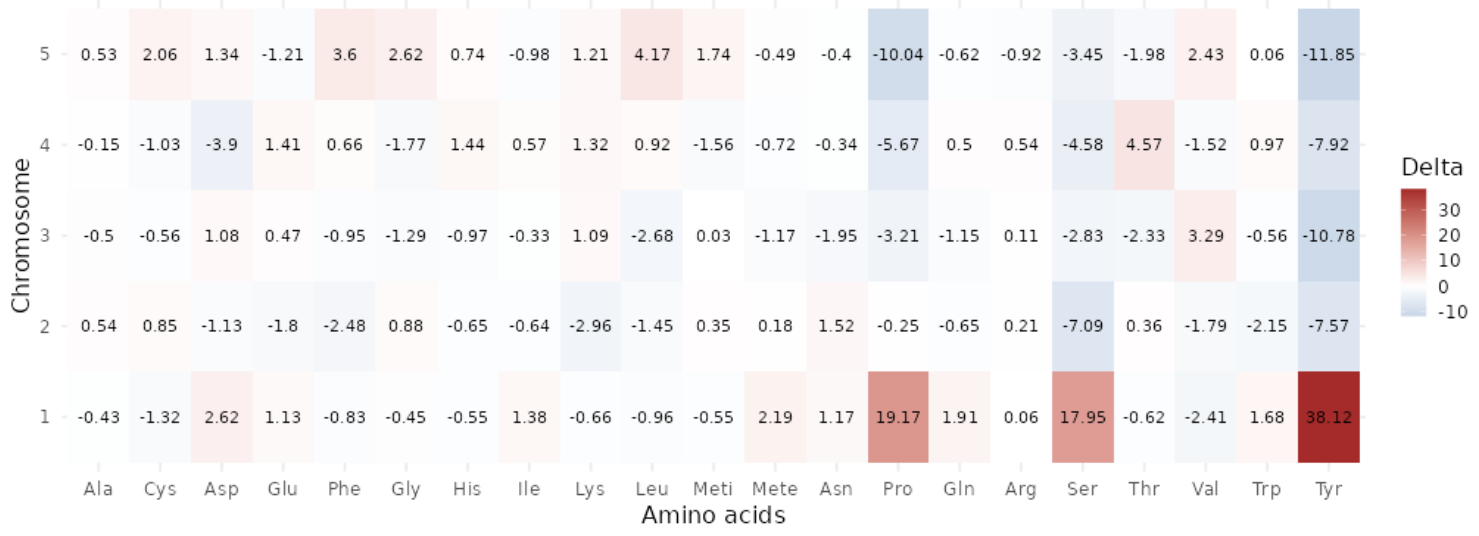

## tRNA Pairwise

Consecutive tRNA pair frequency heatmap

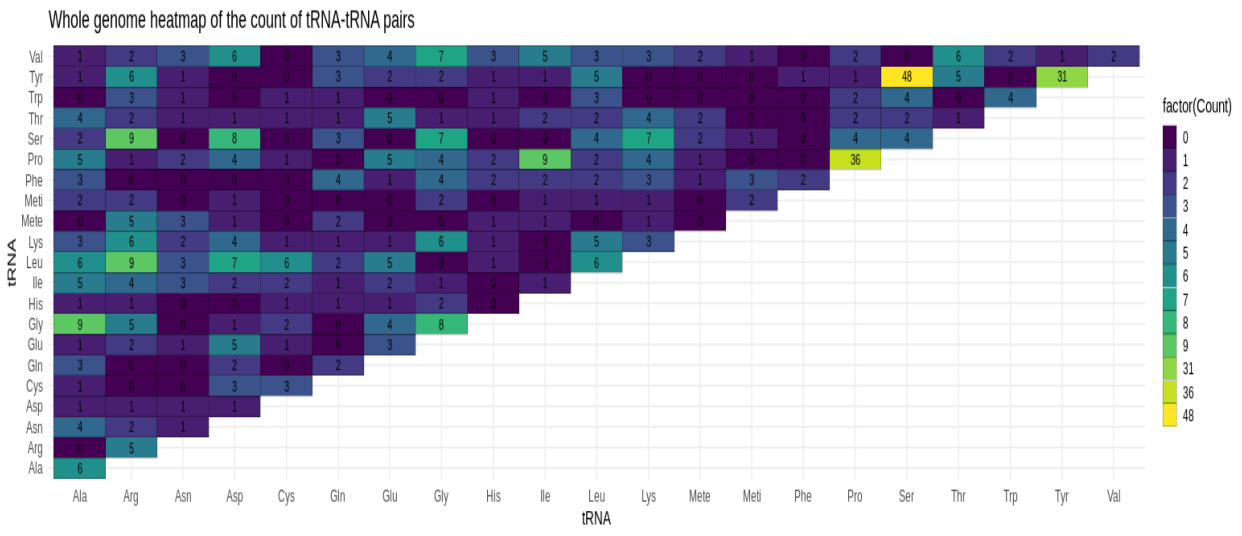

tRNA adjajency distances boxplot

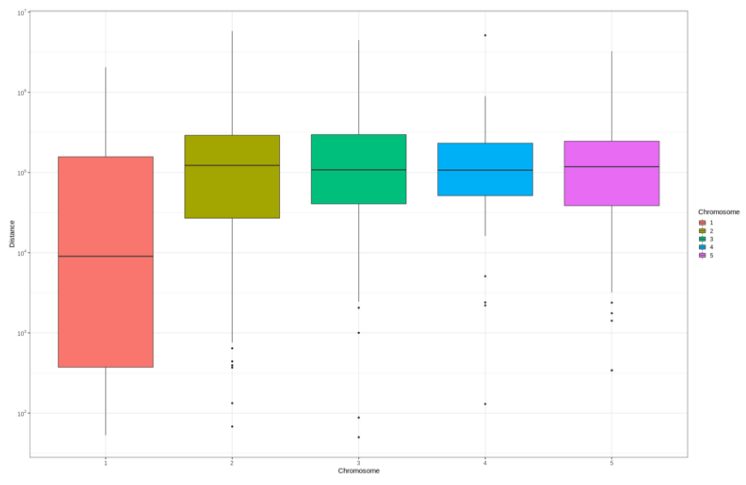

tRNA adjajency distances point

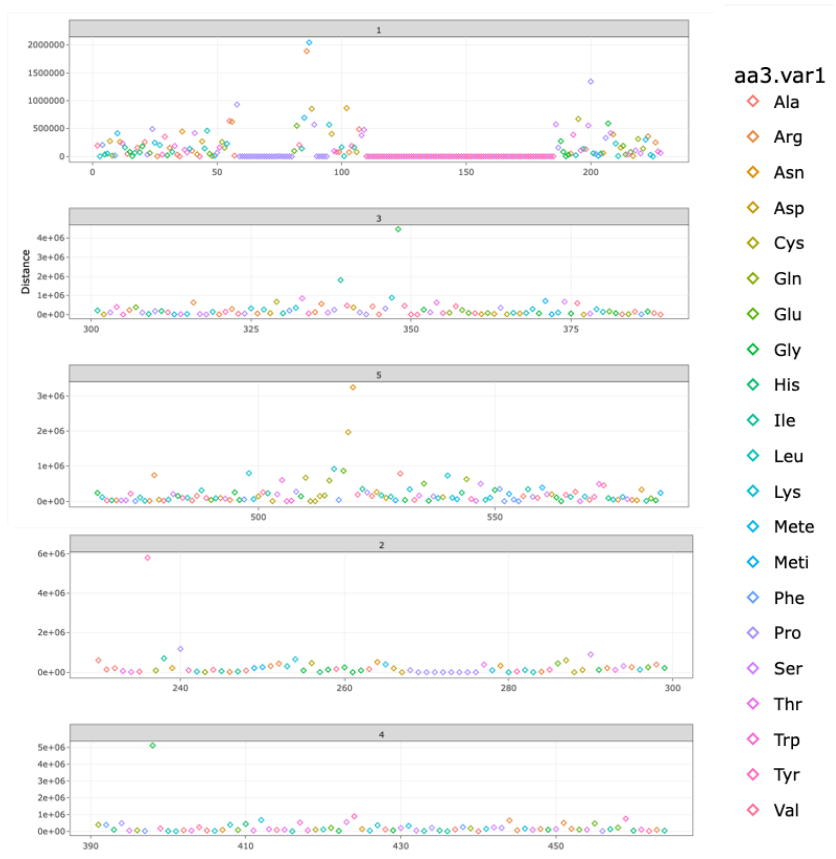

## tRNA Clusters

Cluster plot

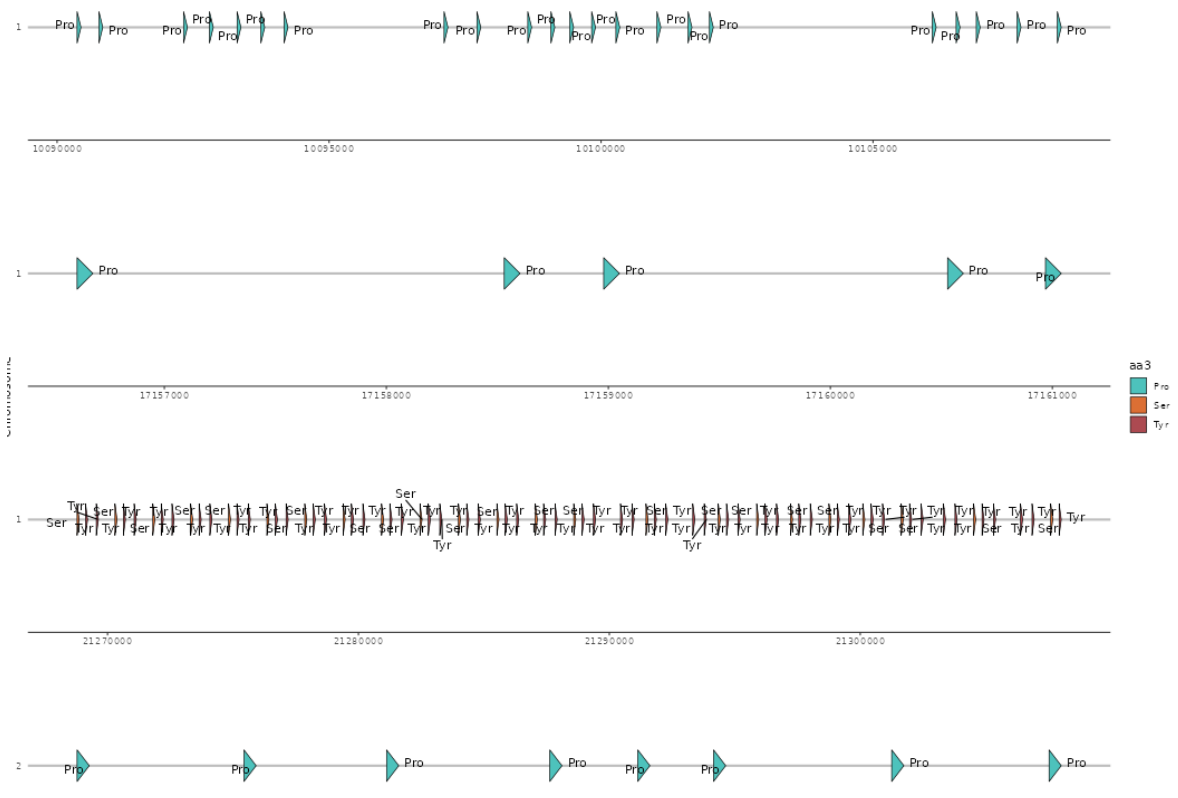

## Centromere Positioning

Chromosome representation with estimated centromere regions

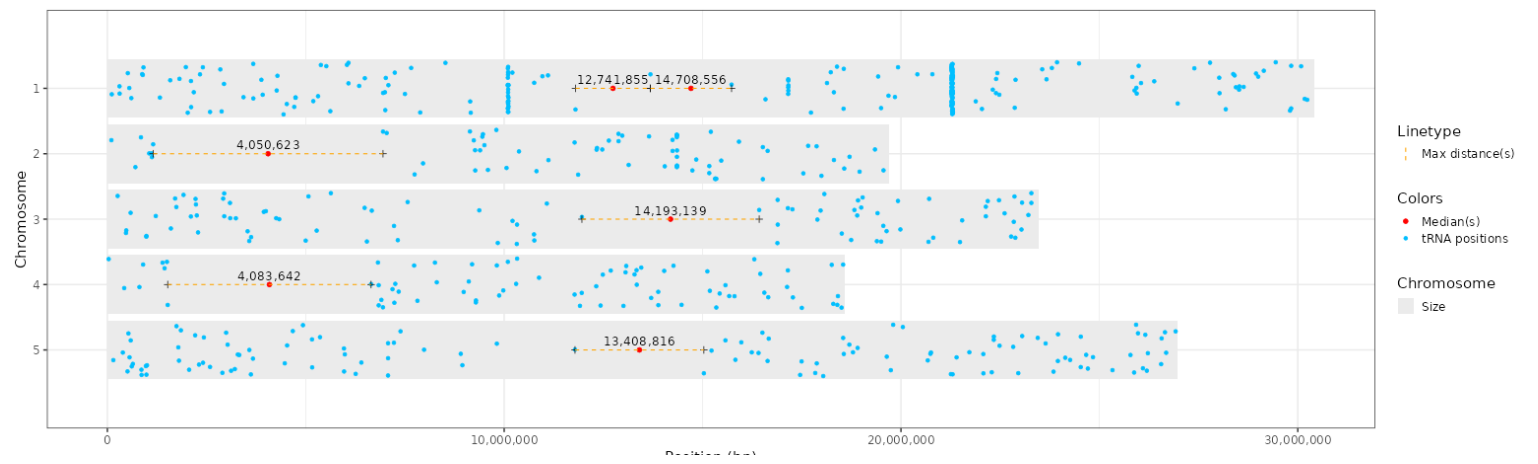

**Supplemental Figure S1.** Data analysis performed on the ShinytRNA platform https://nebula.ibmp.unistra.fr/shinytRNA/ ). Examples of data obtained for Arabidopsis thaliana (Ath).

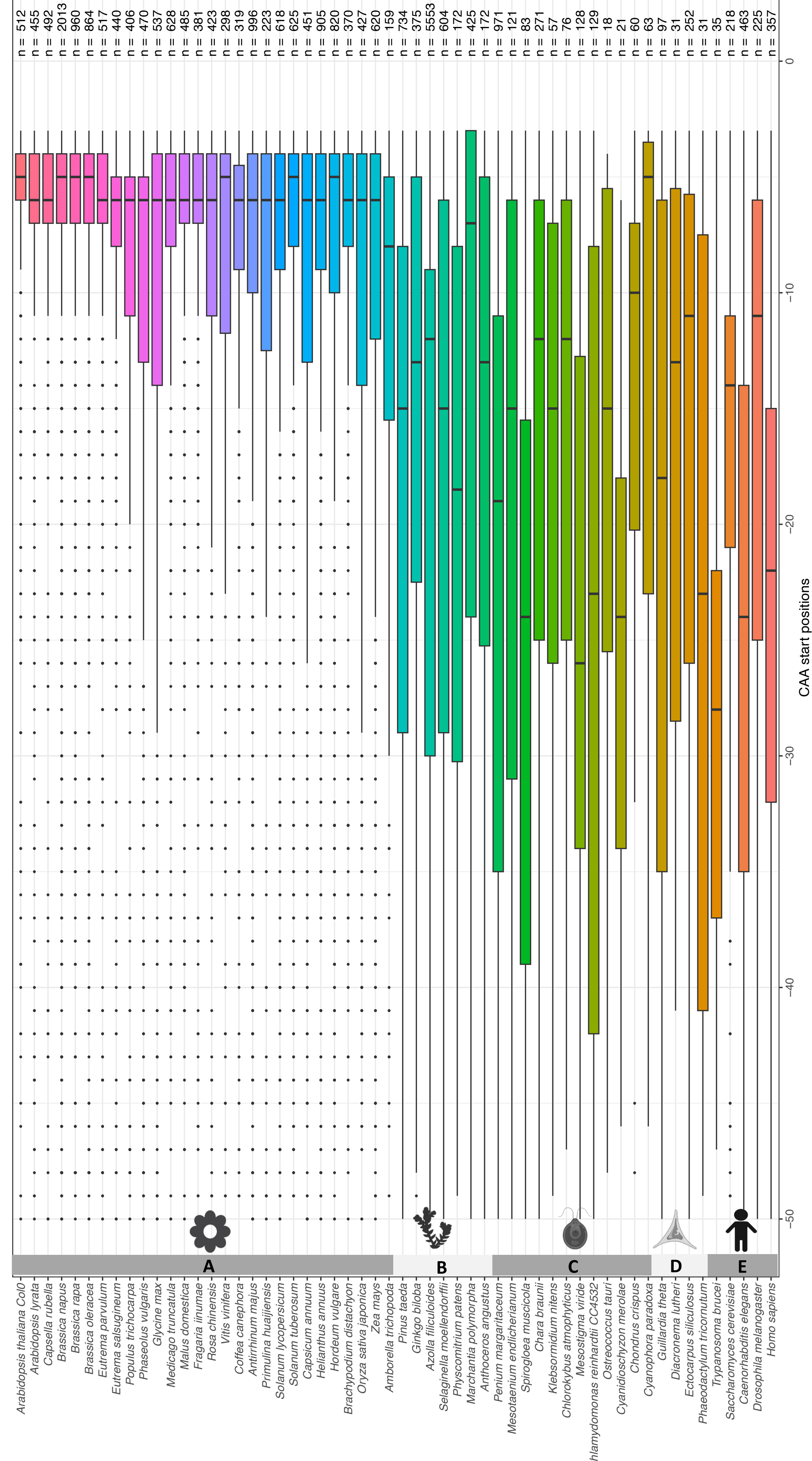

**Figure S2.** Boxplot showing the CAA start position of the first CAA motif in the region upstream of the 5′ end of tDNAs. (**A**) Angiosperms, (**B**) other embryophytes excluding angiosperms, (**C**) other Archaeplastida, (**D**) photosynthetic organisms with secondary plastids, and (**E**) non-photosynthetic organisms. n = number of analyzed tDNAs.

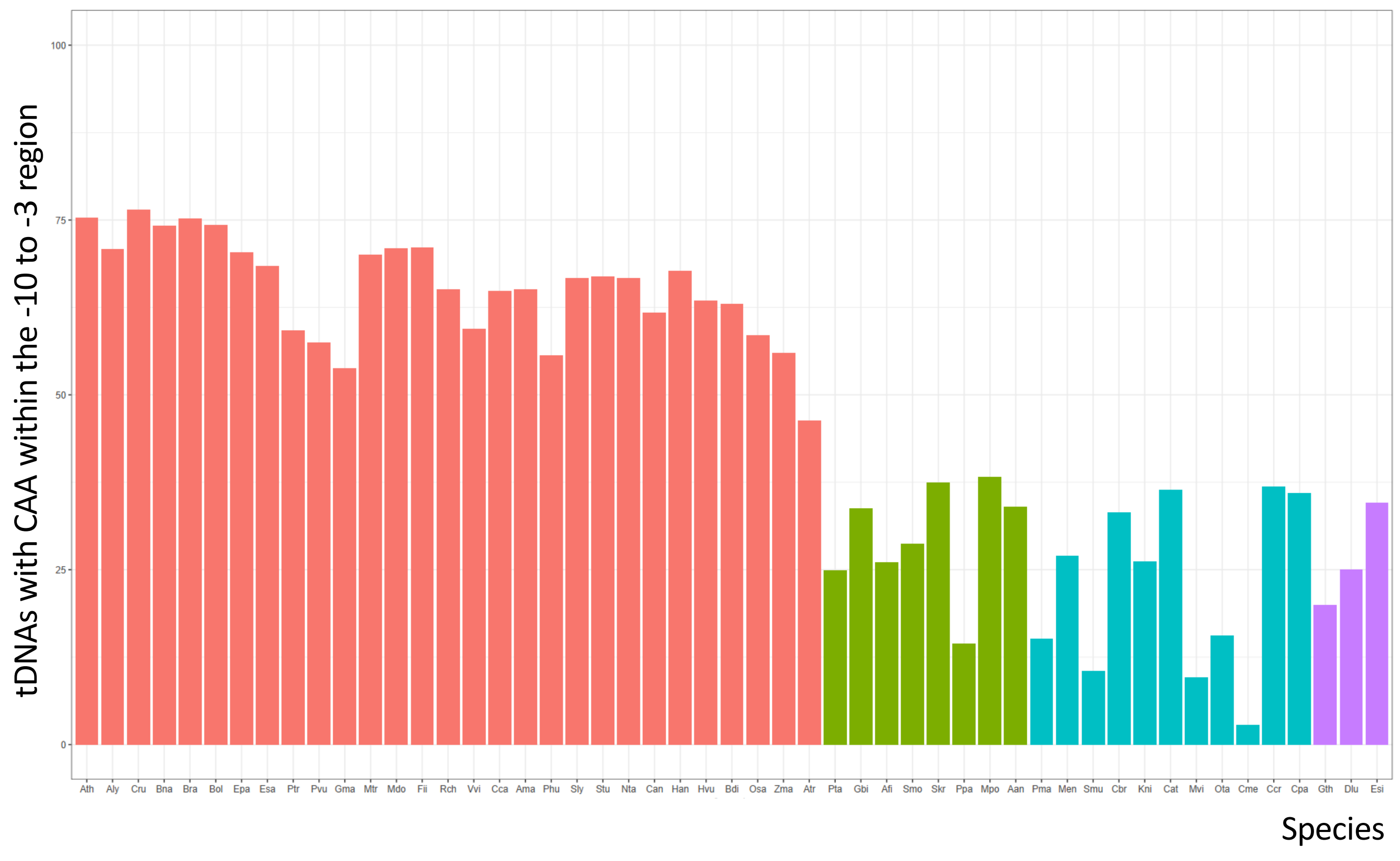

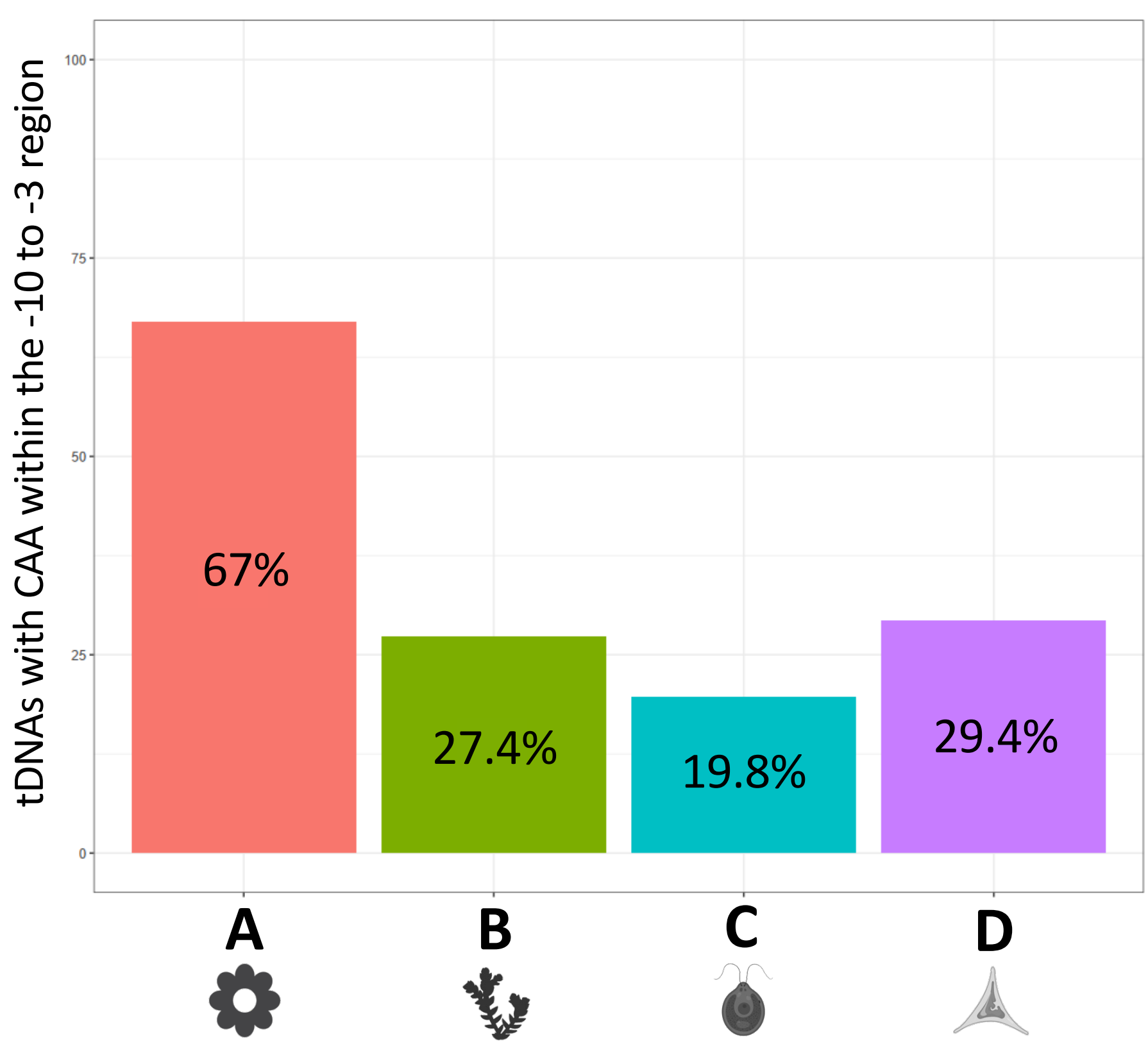

**Figure S3.** Histograms showing the proportion of tDNAs with a CAA motif within the 10 to -3 upstream region in photosynthetic species (**a**), in major photosynthetic groups (**b**). (**A**) Angiosperms, (**B**) other embryophytes excluding angiosperms, (**C**) other Archaeplastida, (**D**) photosynthetic organisms with secondary plastids.

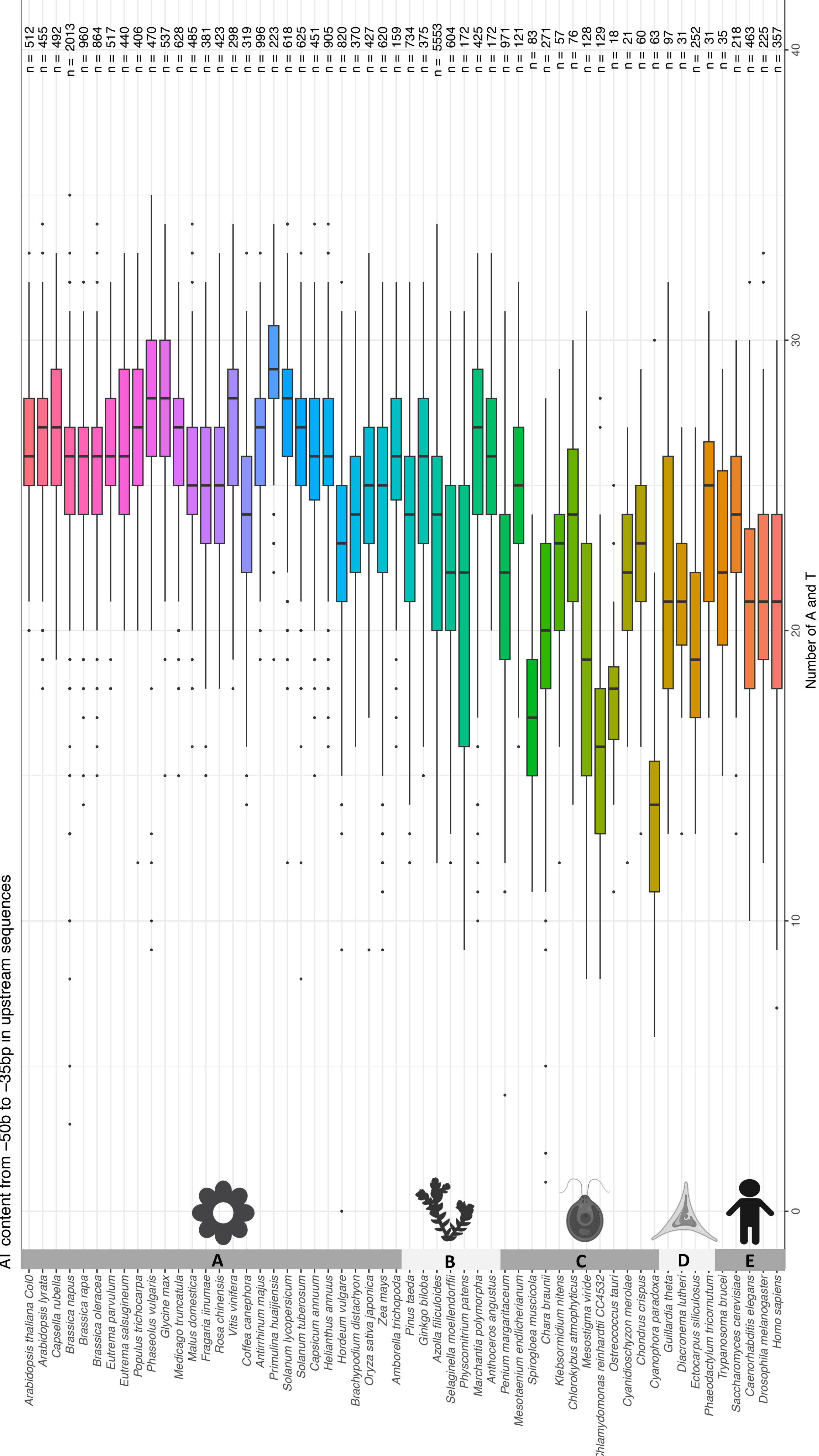

**Figure S4.** Boxplot showing the AT content in the region from –50 to –35 bp upstream of the 5′ end of tDNAs. (**A**) Angiosperms, (**B**) other embryophytes excluding angiosperms, (**C**) other Archaeplastida, (**D**) photosynthetic organisms with secondary plastids, and (**E**) non-photosynthetic organisms.  n = number of analyzed tDNAs.

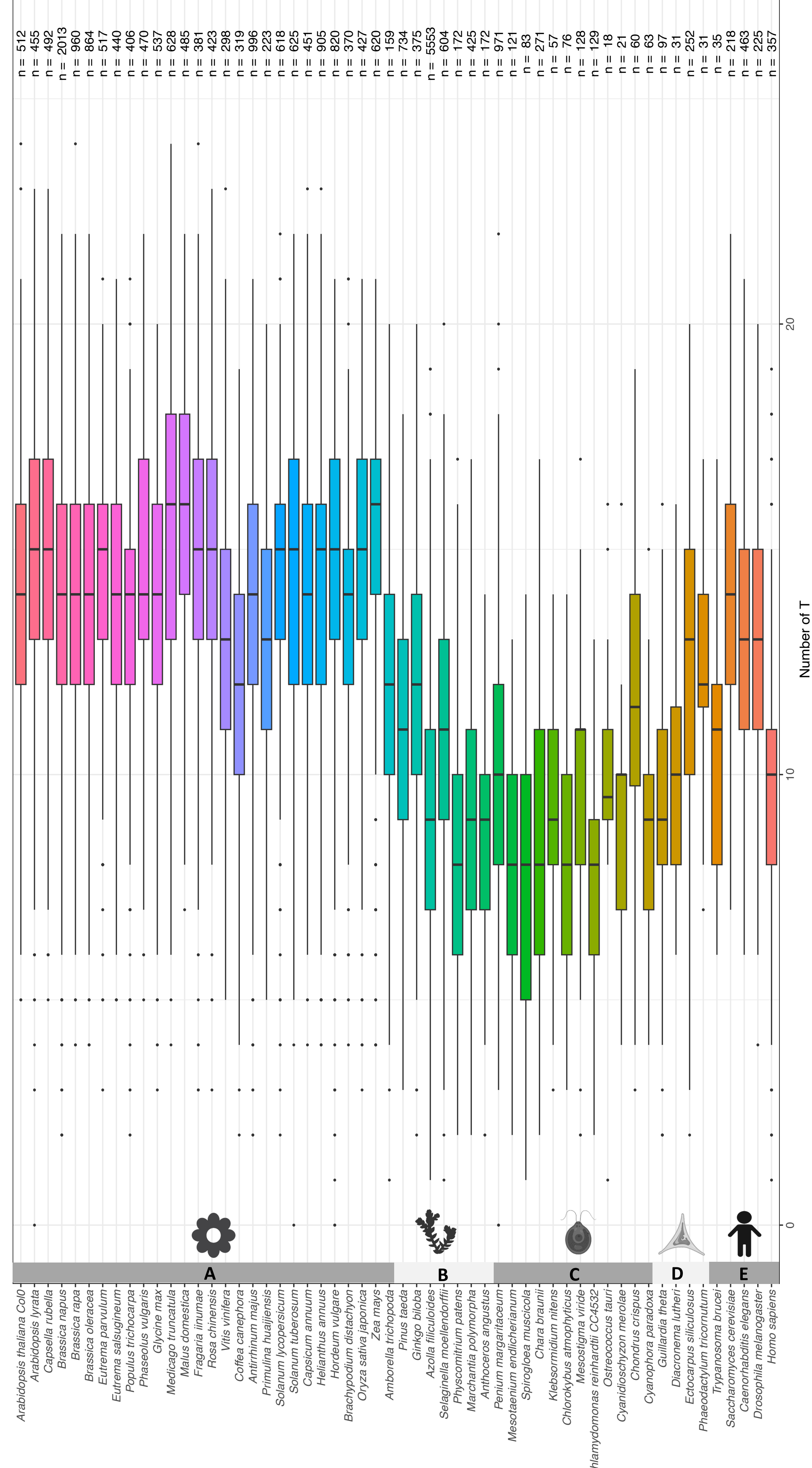

**Figure S5.** Boxplot showing the T content within 25 bp downstream of the 3′ end of tDNAs. (**A**) Angiosperms, (**B**) other embryophytes excluding angiosperms, (**C**) other Archaeplastida, (**D**) photosynthetic organisms with secondary plastids, and (**E**) non-photosynthetic organisms. n = number of analyzed tDNAs.

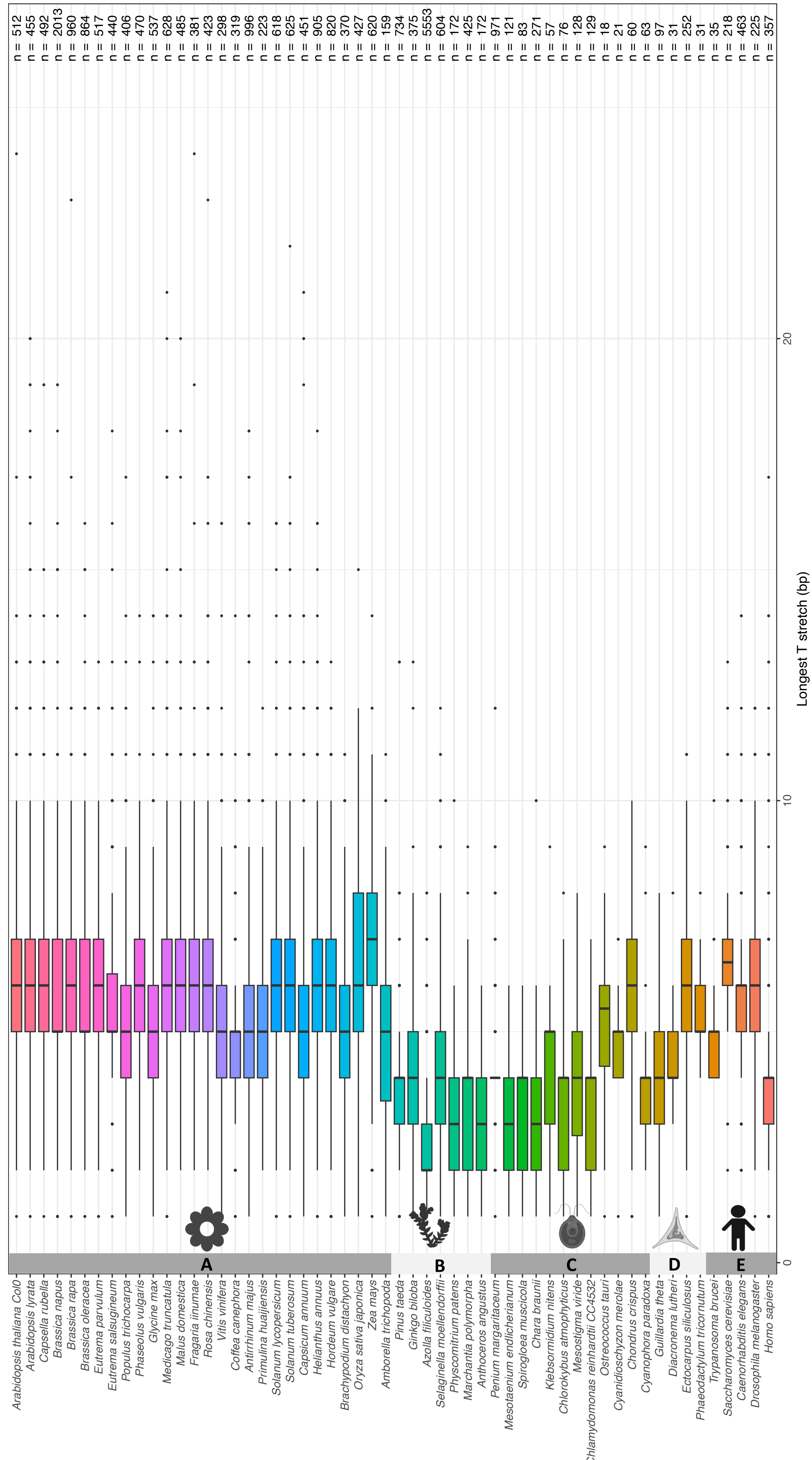

**Figure S6.** Boxplot showing the length of the longest T stretch located within 25 bp downstream of the 3' end of tDNAs. (**A**) Angiosperms, (**B**) other embryophytes excluding angiosperms, (**C**) other Archaeplastida, (**D**) photosynthetic organisms with secondary plastids, and (**E**) non-photosynthetic organisms. n = number of analyzed tDNAs.

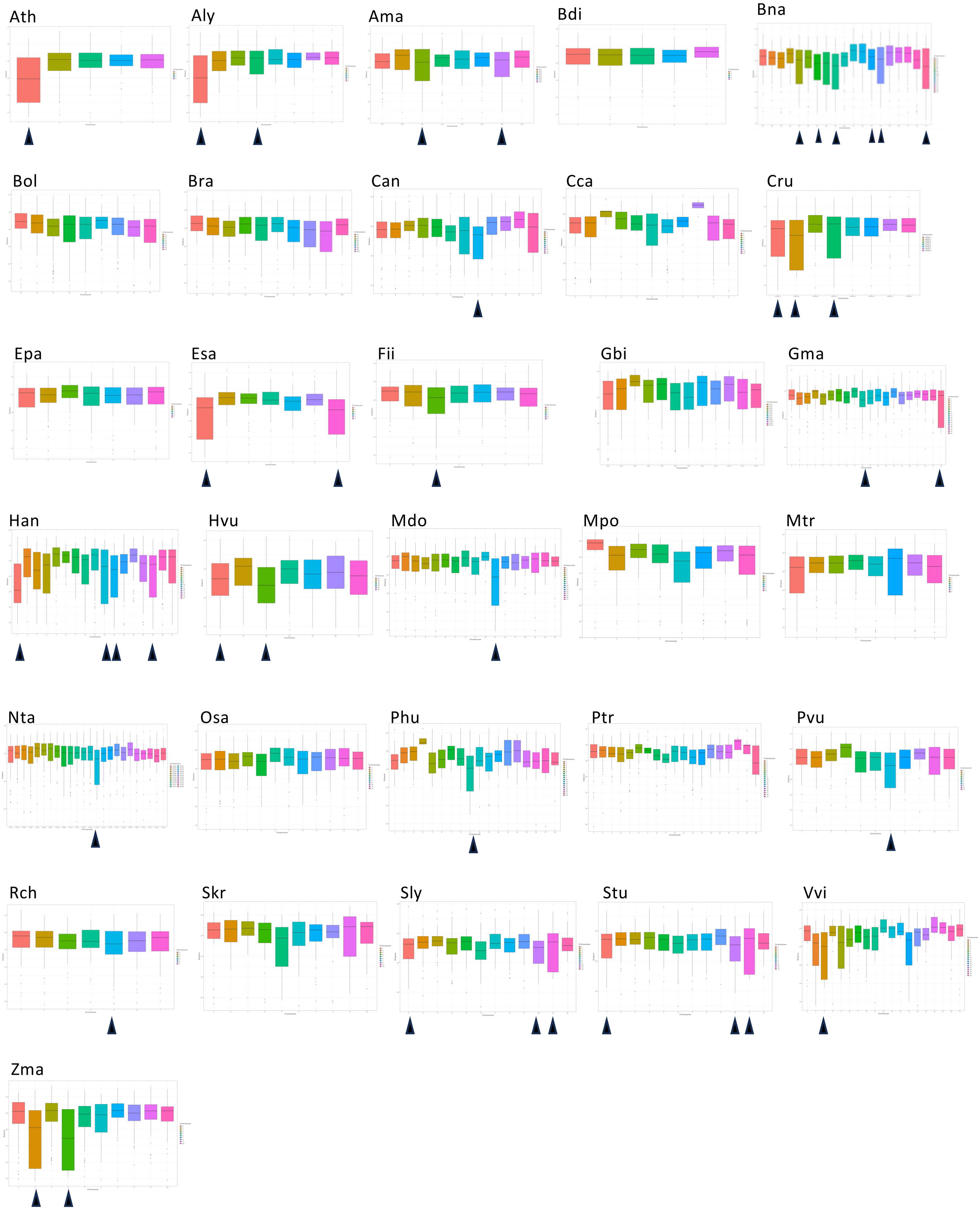

**Figure S7.** Boxplots showing the distance (d) between consecutive tDNAs on chromosomes from land plant species. Arrow heads indicate the main clusters on the respective chromosomes. Names of organisms are abbreviated as in Table S1. All boxplots are available at https://nebula.ibmp.unistra.fr/shinytRNA/).

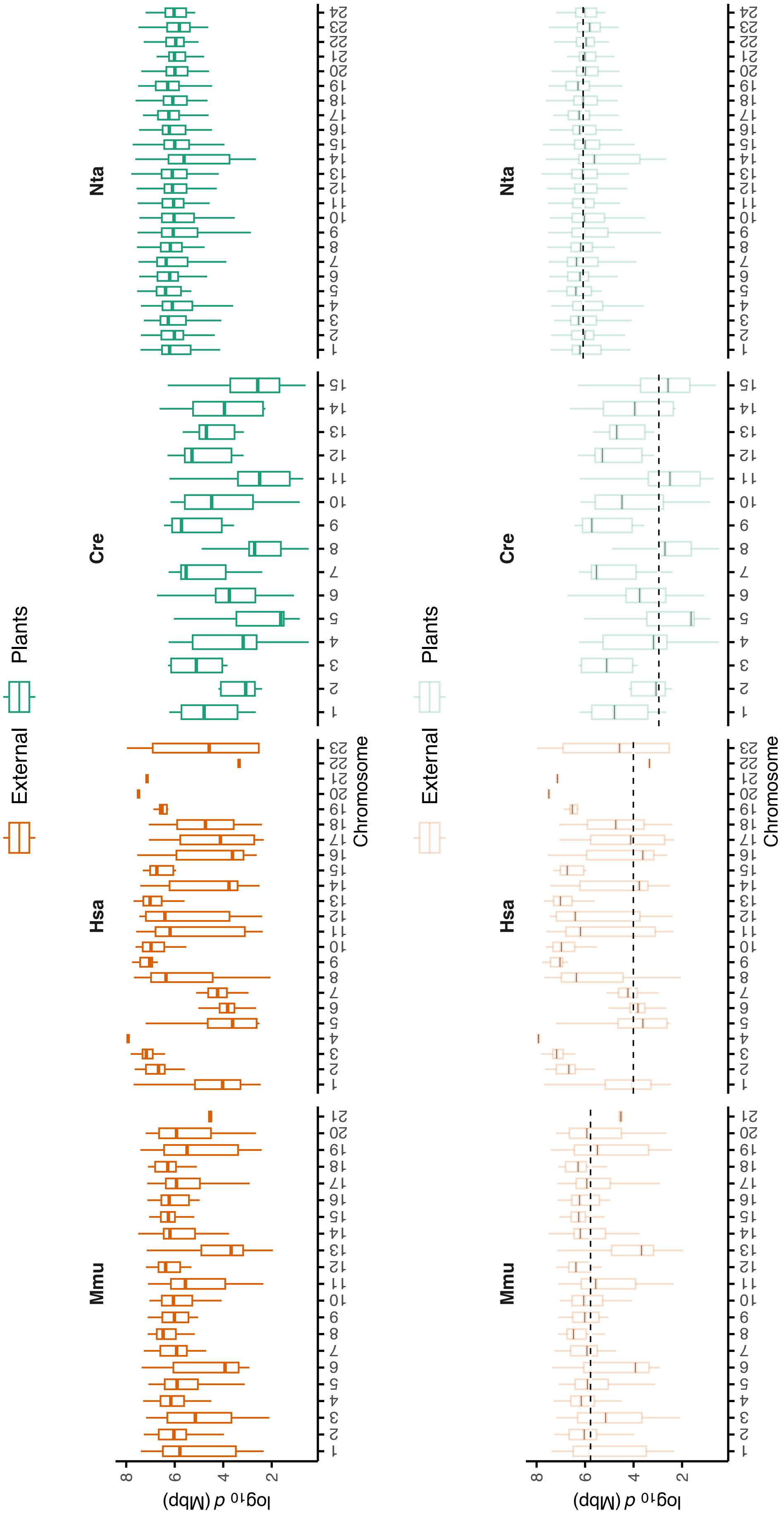

**Figure S8:** Comparison of tDNA organization in *Chlamydomonas reinhardtii* versus a representative plant species (*Nicotiana tabacum*) and two non-plant species (*Mus musculus* and *Homo sapiens*). Boxplots display the distribution of intrachromosomal intergenic distances between tDNAs ($\log_{10}$ *d*, in Mbp). Horizontal dashed lines indicate median distances.

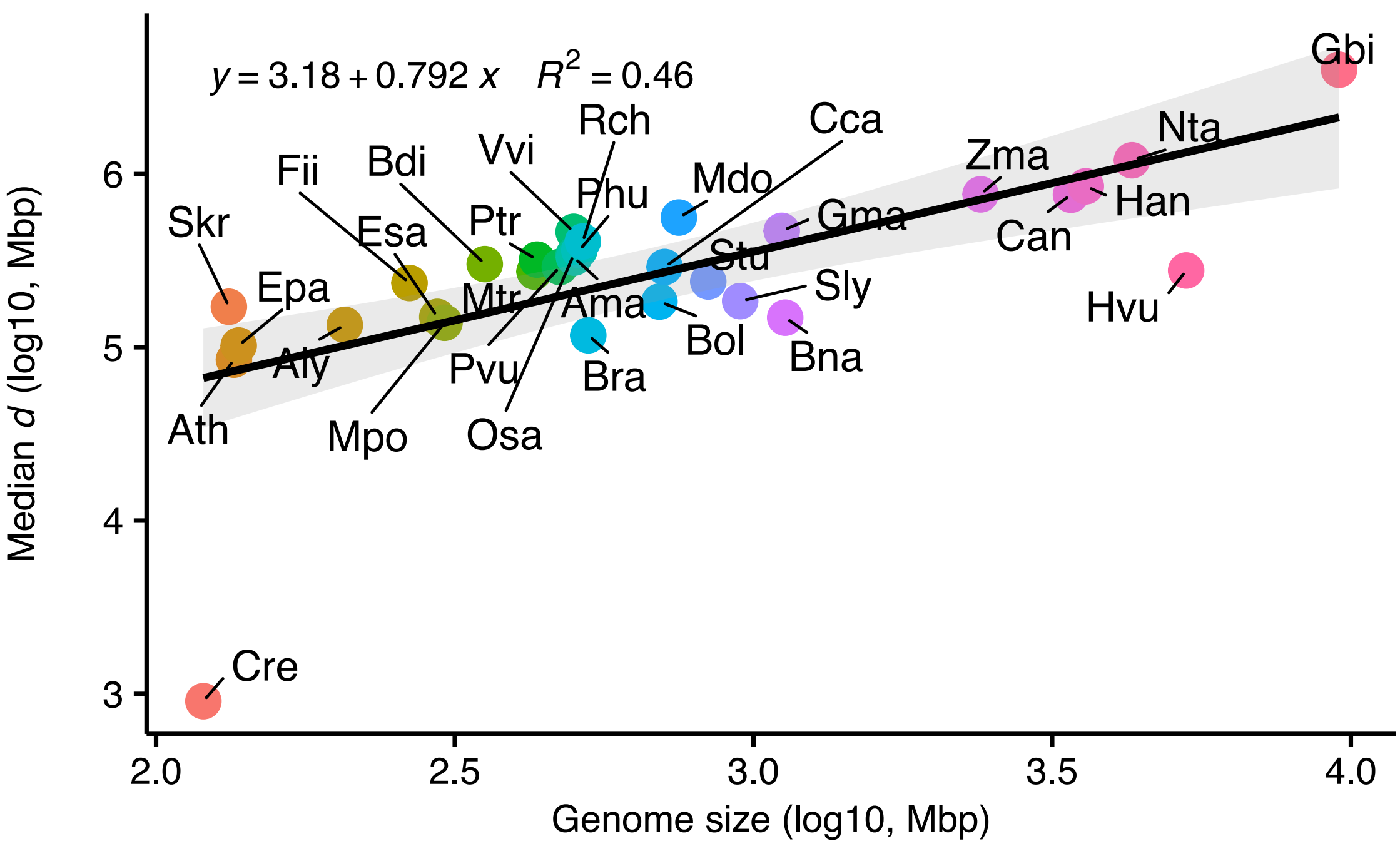

**Figure S9**. tDNA organization in the green lineage including *Chlamydomonas reinhardtii*. The representation is identical to Figure 5b but including *C. reinhardtii* (Cre).

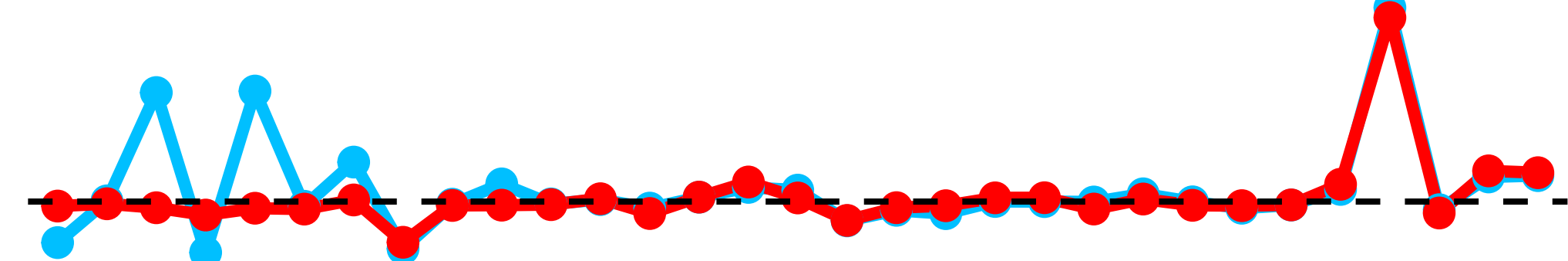

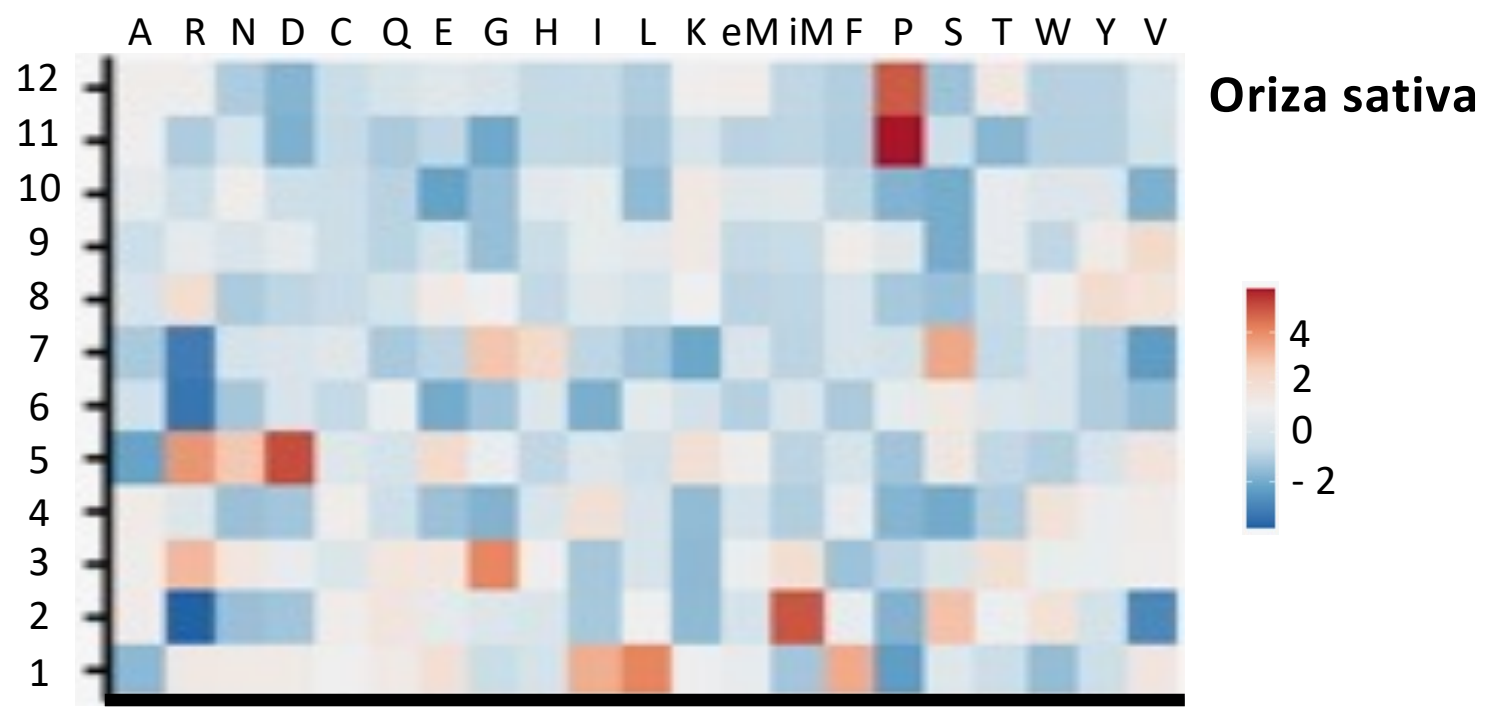

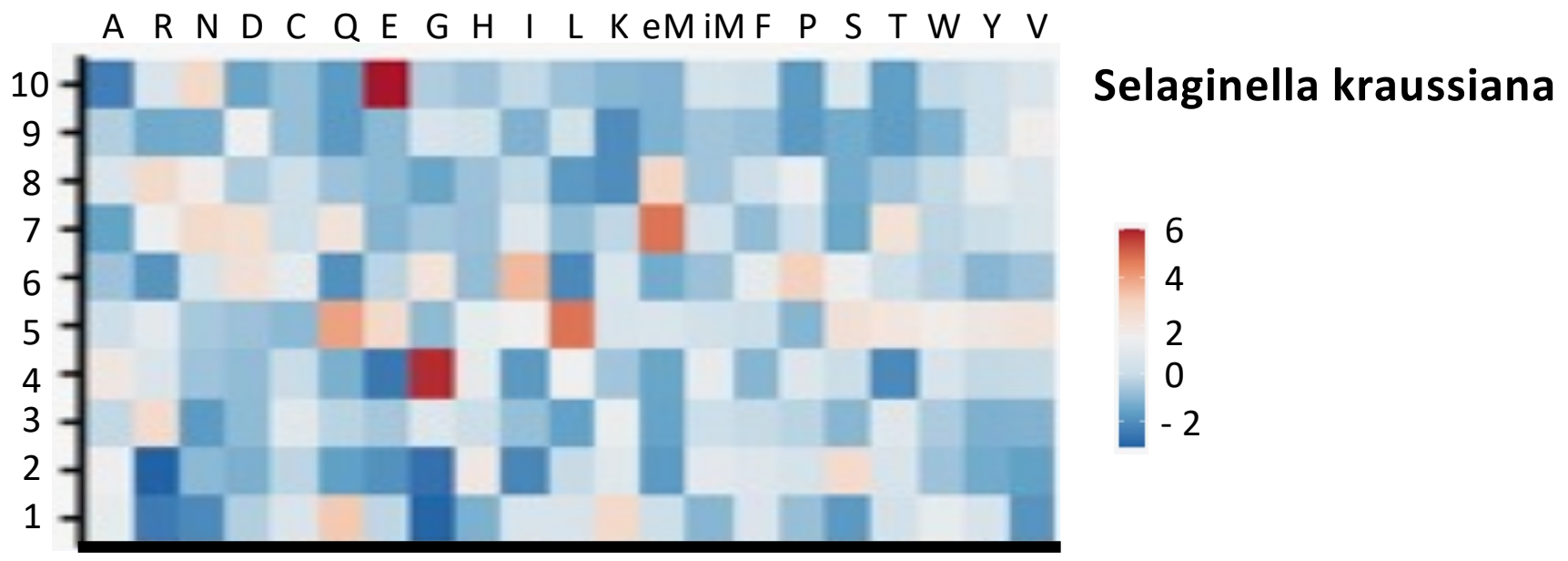

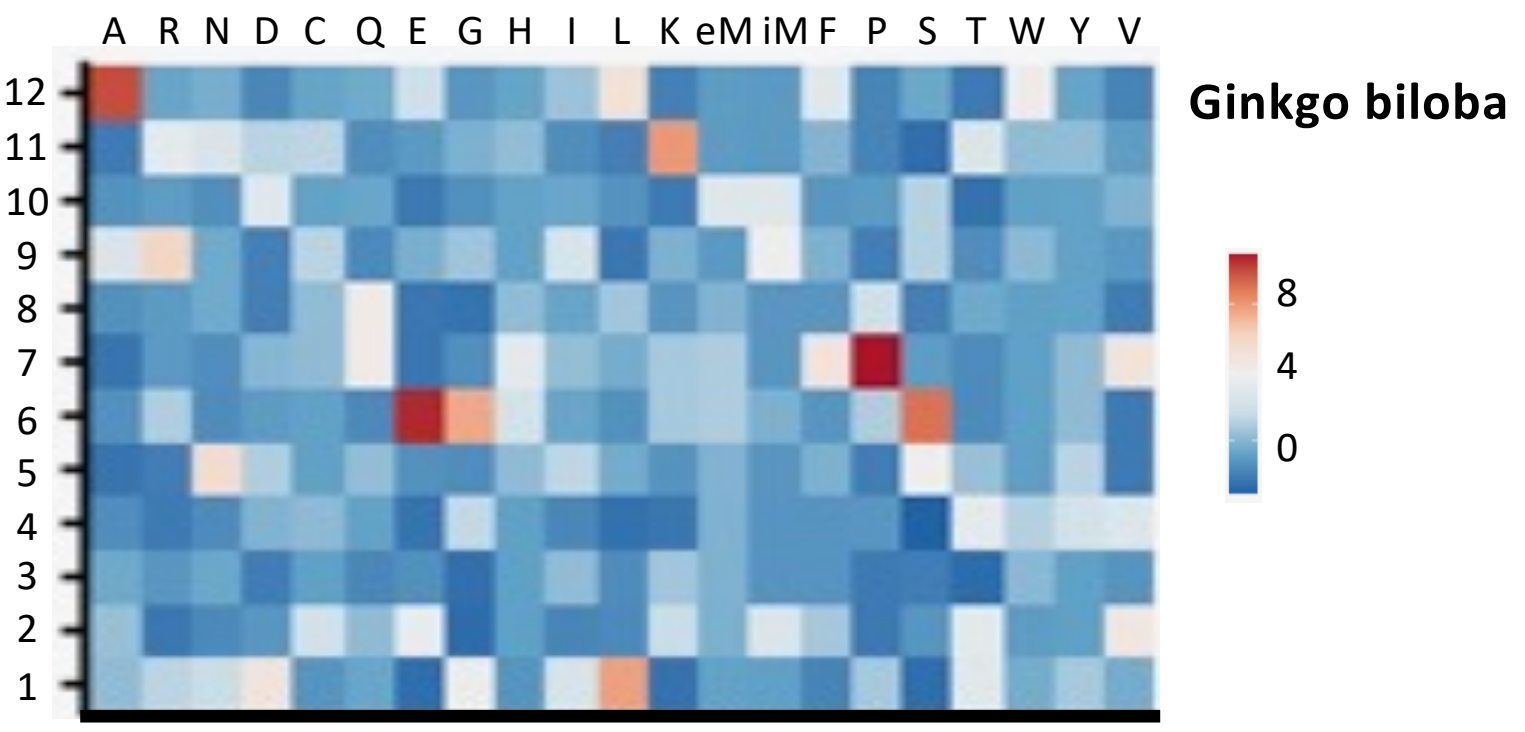

**Supplemental Figure S10.** Heatmaps showing the chromosomal distribution of isoacceptor tDNA families in three species. Rows correspond to chromosomes, columns to amino acid families (single-letter code). Colors indicate deviation from the expected random distribution normalized by genome size, with red denoting enrichment and blue depletion.

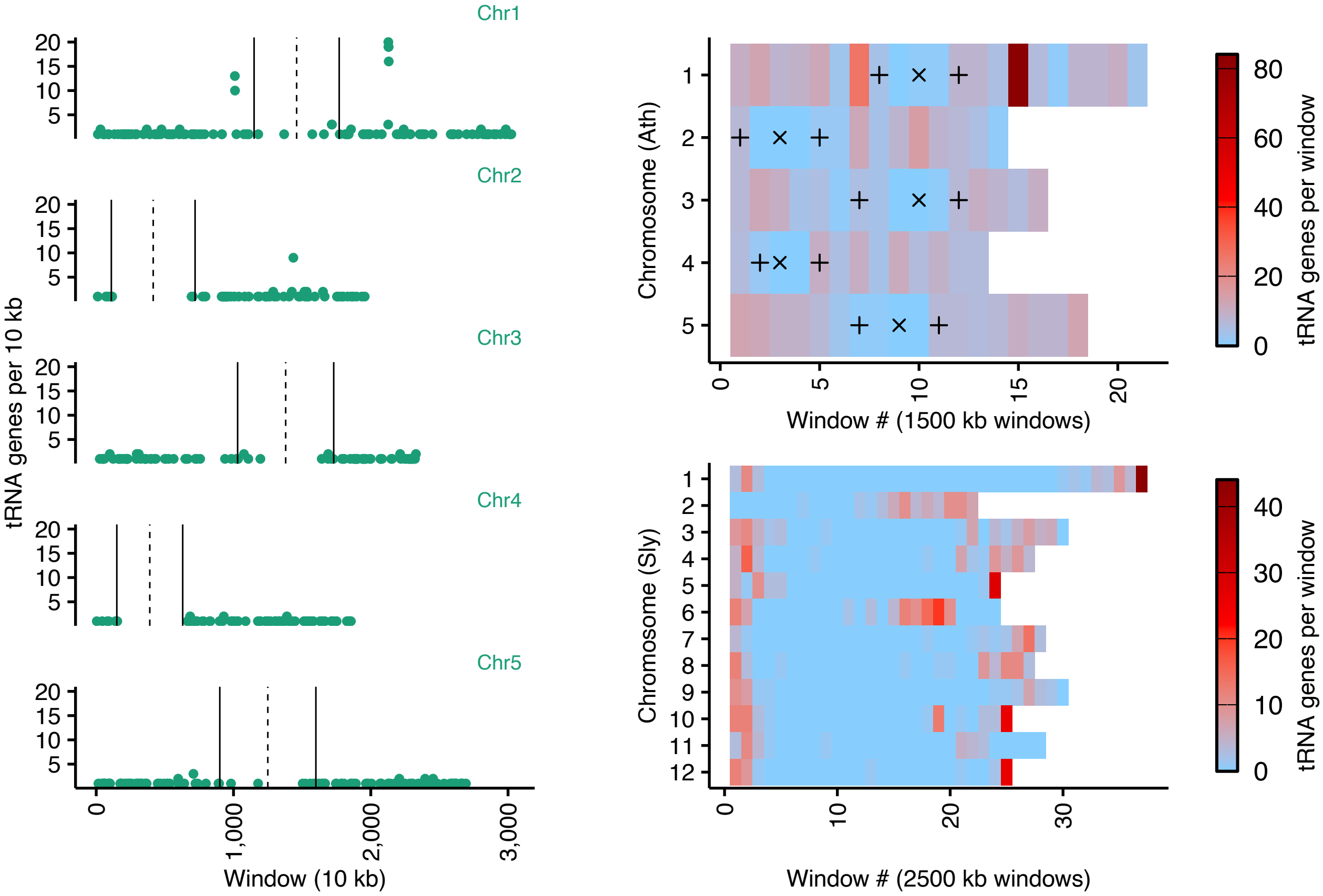

**Supplemental Figure S11.** Chromosomal distribution of tDNAs reveals large-scale genomic organization in plants. **A**) Number of tDNAs within consecutive 10-kb windows along the chromosomes of *Arabidopsis thaliana*. Solid and dotted vertical black lines indicate the boundaries and central coordinates of centromeric regions, respectively. **B**) Heatmap of tDNA density across the chromosomes of *A. thaliana* and *Solanum lycopersicum*. The (+) and (×) symbols mark the approximate centromeric regions and their central coordinates in *A. thaliana*, respectively.+--

Minimum gene number

3 5 10

Maximum genomic distance

1K 5K 50K

Clustering ratio (%)

0 20 40 60 80 100

**Supplemental Figure S12.** tDNA cluster identification upon different filtering conditions. White indicates no clustering (all genes isolated), while dark red indicates full clustering (all genes grouped). In *Cyanydoschyzon merolae* (Cme) and *Phaelodactylum tricornutum* (Phatr), asparagine (D) and cysteine (C) tDNAs could not be confidently annotated and were therefore excluded (black crossed-out regions). Species labels identical to Table S1.

## Minimum 1 $\text{tRNA}^{\text{Pro}}$ gene, filtering parameters: min($n$) = 5 and max($d$) = 5 kb

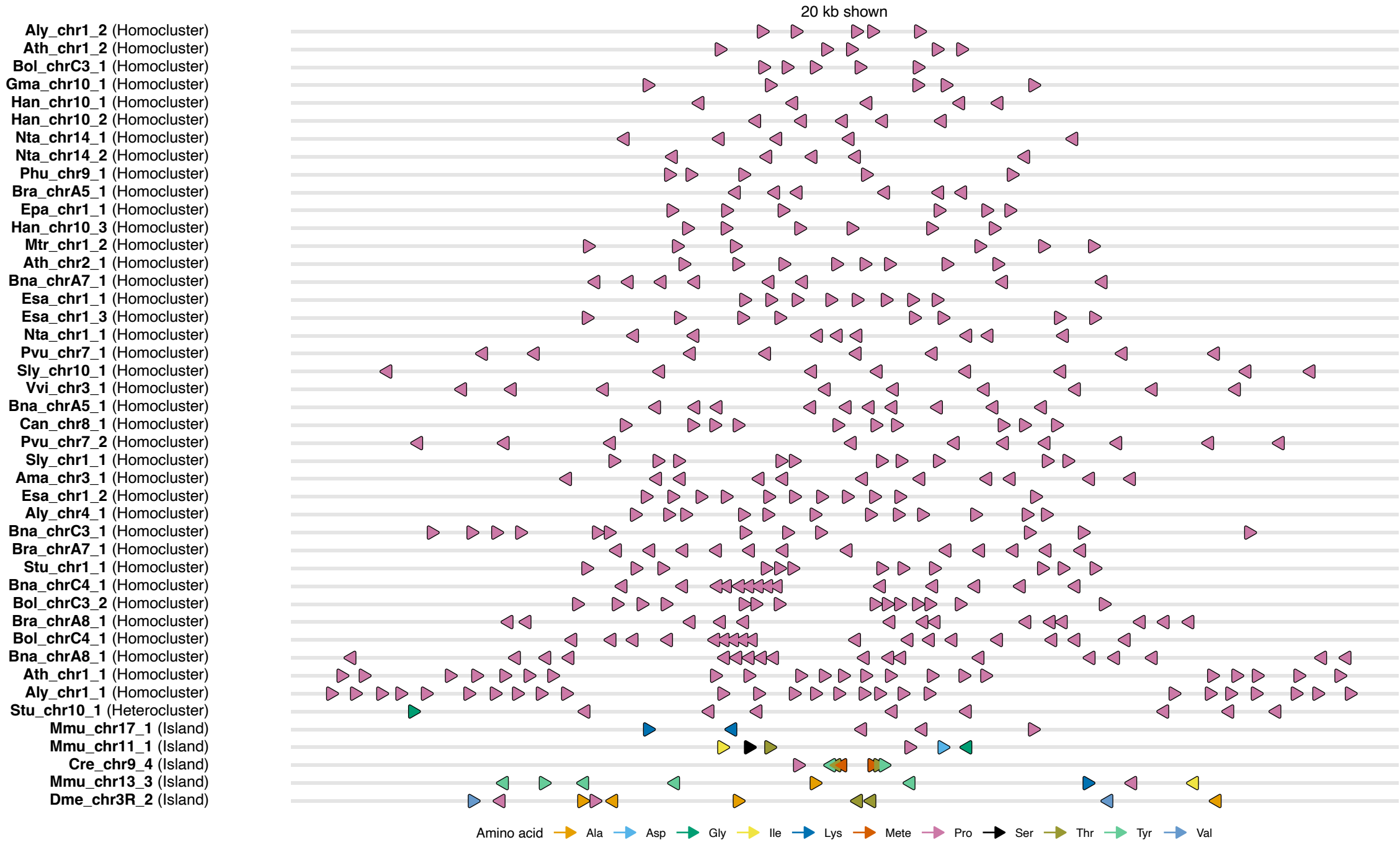

## Minimum 1 $\text{tRNA}^{\text{Ser}}$ gene, filtering parameters: min($n$) = 5 and max($d$) = 5 kb

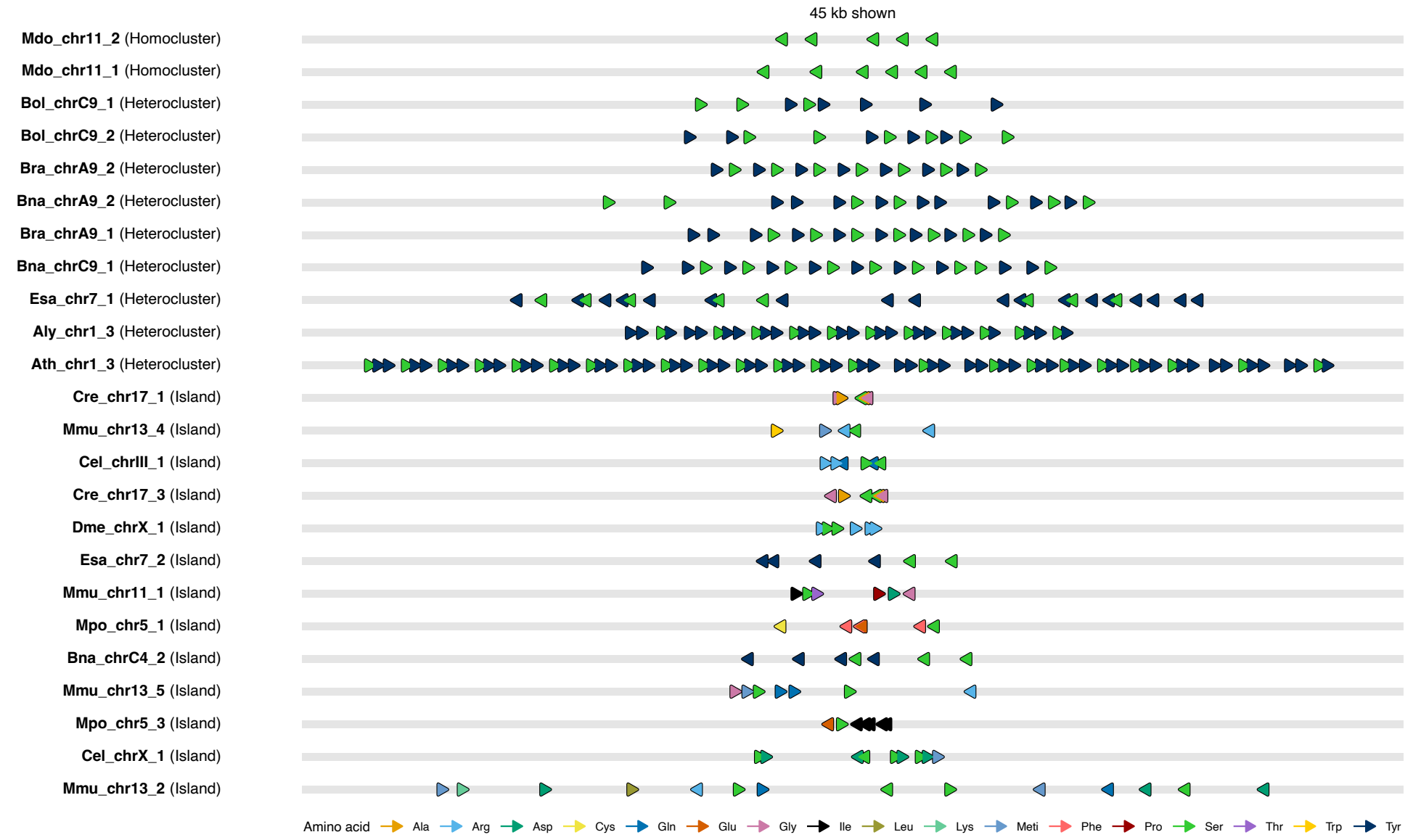

## Minimum 1 $\text{tRNA}^{\text{Tyr}}$ gene, filtering parameters: min($n$) = 5 and max($d$) = 5 kb

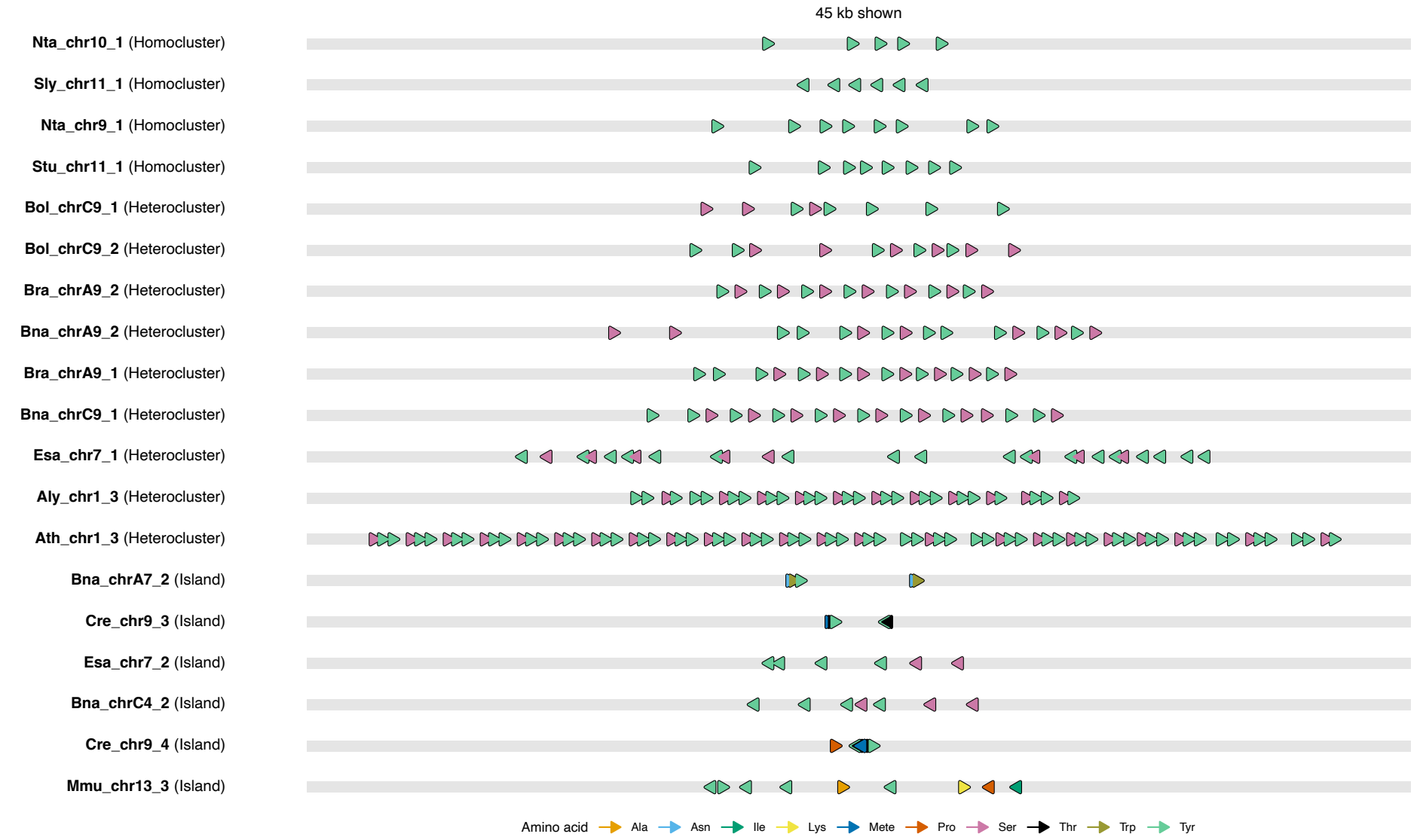

**Supplemental Figure S13.** Identification of proline-, serine-, and tyrosine-containing tDNA clusters.

**Table S1**: Photosynthetic and non-phosynthetic species analyzed in this study.

| Major group | Clade | Abbreviation | Species | genome size (Mb) | Nuclear tRNA gene number |
|---|---|---|---|---|---|
| Angiospserm (A) | Angiosperm | Ath | *Arabidopsis thaliana* | 135 | 585 |
| | Angiosperm | Aly | *Arabidopsis lyrata* | 206,8 | 546 |
| | Angiosperm | Cru | *Capsella rubella* | 133,5 | 570 |
| | Angiosperm | Bna | *Brassica napus* | 976,2 | 2201 |
| | Angiosperm | Bra | *Brassica rapa* | 353,2 | 1047 |
| | Angiosperm | Bol | *Brassica oleracea* | 488,9 | 962 |
| | Angiosperm | Epa | *Eutrema parvulum* | 137,5 | 585 |
| | Angiosperm | Esa | *Eutrema salsigineum* | 295,6 | 500 |
| | Angiosperm | Ptr | *Populus trichocarpa* | 434,3 | 505 |
| | Angiosperm | Pvu | *Phaseolus vulgaris* | 521,2 | 591 |
| | Angiosperm | Gma | *Glycine max* | 978,9 | 700 |
| | Angiosperm | Mtr | *Medicago truncatula* | 430 | 704 |
| | Angiosperm | Mdo | *Malus domestica* | 703,7 | 562 |
| | Angiosperm | Fii | *Fragaria iinumae* | 204,7 | 428 |
| | Angiosperm | Rch | *Rosa chinensis* | 513,8 | 481 |
| | Angiosperm | Vvi | *Vitis vinifera* | 486,2 | 367 |
| | Angiosperm | Cca | *Coffea canephora* | 568,8 | 378 |
| | Angiosperm | Ama | *Antirrhinum majus* | 500,8 | 578 |
| | Angiosperm | Phu | *Primulina huijiensis* | 470 | 293 |
| | Angiosperm | Sly | *Solanum lycopersicum* | 828,3 | 722 |
| | Angiosperm | Stu | *Solanum tuberosum* | 706 | 736 |
| | Angiosperm | Nta | *Nicotiana tabacum* | 3990 | 1228 |
| | Angiosperm | Can | *Capsicum annuum* | 2935,9 | 526 |
| | Angiosperm | Han | *Helianthus annuus* | 3027,8 | 1040 |
| | Angiosperm | Hvu | *Hordeum vulgare* | 4834,4 | 985 |
| | Angiosperm | Bdi | *Brachypodium distachyon* | 271,3 | 462 |
| | Angiosperm | Osa | *Oryza sativa* | 374,4 | 504 |
| | Angiosperm | Zma | *Zea mays* | 2135,1 | 813 |
| | Angiosperm | Atr | *Amborella trichopoda* | 71 | 207 |
| Non-angiosperm embryophyte (B) | Gymnosperm | Pta | *Pinus taeda* | 22105 | 1012 |
| | Gymnosperm | Gbi | *Ginkgo biloba* | 9555,4 | 506 |
| | Fern | Afi | *Azolla filiculoides* | 622,7 | 7496 |
| | Lycophyte | Smo | *Selaginella moellendorffii* | 213,2 | 920 |
| | Lycophyte | Skr | *Selaginella kraussiana* | 114,5 | 456 |
| | Mosse | Ppa | *Physcomitrium patens* | 480,2 | 382 |
| | Liverwort | Mpo | *Marchantia polymorpha* | 219 | 611 |
| | Hornwort | Aan | *Anthoceros angustus* | 119,7 | 226 |
| other Archaeplastida (C) | Streptophyte | Pma | *Penium margaritaceum* | 3662 | 1463 |
| | Streptophyte | Men | *Mesotaenium endlicherianum* | 173,9 | 174 |
| | Streptophyte | Smu | *Spirogloea muscicola* | 170,8 | 152 |
| | Streptophyte | Cbr | *Chara braunii* | 1751,5 | 373 |
| | Streptophyte | Kni | *Klebsormidium nitens* | 104,2 | 84 |
| | Streptophyte | Cat | *Chlorokybus atmophyticus* | 74,7 | 96 |
| | Streptophyte | Mvi | *Mesostigma viride* | 441,9 | 249 |
| | Chlorophyte | Cre | *Chlamydomonas reinhardtii* | 111,5 | 324 |
| | Chlorophyte | Ota | *Ostreococcus tauri* | 13 | 45 |
| | Rhodophyte | Cme | *Cyanidioschyzon merolae* | 16,7 | 35 |
| | Rhodophyte | Ccr | *Chondrus crispus* | 105 | 84 |
| | Glaucophyte | Cpa | *Cyanophora paradoxa* | 100 | 103 |
| Secondary endosymbionts (D) | Cryptophyte | Gth | *Guillardia theta* | 87,8 | 160 |
| | Haptophyte | Dlu | *Diacronema lutheri* | 43,6 | 56 |
| | Heterokontophyte | Esi | *Ectocarpus siliculosus* | 210,4 | 341 |
| | Heterokontophyte | Ptri | *Phaeodactylum tricornutum* | 19,7 | 52 |
| Non-plant organisms (E | Metazoa | Hsa | *Homo sapiens* | 3300 | 429 |
| | Metazoa | Mmu | *Mus musculus* | 2800 | 400 |
| | Metazoa | Cel | *Caenorhabditis elegans* | 100 | 568 |
| | Metazoa | Dme | *Drosophila melanogaster* | 180 | 290 |
| | Fungi | Sce | *Saccharomyces cerevisiae* | 12 | 275 |
| | Amoebozoa | Ddi | *Dictyostelium discoidum* | 34 | 403 |
| | Discoba | Tbr | *Trypanosoma brucei* | 26 | 64 |

**Table S2: All Pairwise Wilcoxon Rank Sum Tests with Benjamini & Hochberg (BH) adjustement**

**T counts**

| Group1 | Group2 | p.value | signif |
|---|---|---|---|
| Ground | Flowers | 0 | *** |
| Water | Flowers | 0 | *** |
| Photo | Flowers | 1,98E-55 | *** |
| Misc | Flowers | 2,98E-108 | *** |
| Water | Ground | 0,03177077 | * |
| Photo | Ground | 8,71E-45 | *** |
| Misc | Ground | 2,59E-193 | *** |
| Photo | Water | 6,94E-34 | *** |
| Misc | Water | 3,19E-121 | *** |
| Misc | Photo | 0,00178902 | ** |

**Longest T stretch**

| Group1 | Group2 | p.value | signif |
|---|---|---|---|
| Ground | Flowers | 0 | *** |
| Water | Flowers | 0 | *** |
| Photo | Flowers | 3,61E-07 | *** |
| Misc | Flowers | 8,95E-11 | *** |
| Water | Ground | 3,51E-197 | *** |
| Photo | Ground | 1,68E-143 | *** |
| Misc | Ground | 0 | *** |
| Photo | Water | 2,17E-60 | *** |
| Misc | Water | 8,71E-147 | *** |
| Misc | Photo | 0,25355499 | |

**Longest CAA start**

| Group1 | Group2 | p.value | signif |
|---|---|---|---|
| Ground | Flowers | 0 | *** |
| Water | Flowers | 0 | *** |
| Photo | Flowers | 3,55E-42 | *** |
| Misc | Flowers | 0 | *** |
| Water | Ground | 7,42E-24 | *** |
| Photo | Ground | 0,6442986 | |
| Misc | Ground | 7,38E-39 | *** |
| Photo | Water | 0,00032036 | *** |
| Misc | Water | 0,00508479 | ** |
| Misc | Photo | 1,26E-08 | *** |

**AT counts**

| Group1 | Group2 | p.value | signif |
|---|---|---|---|
| Ground | Flowers | 0 | *** |
| Water | Flowers | 0 | *** |
| Photo | Flowers | 3,68E-132 | *** |
| Misc | Flowers | 0 | *** |
| Water | Ground | 3,42E-133 | *** |
| Photo | Ground | 6,81E-45 | *** |
| Misc | Ground | 1,37E-65 | *** |
| Photo | Water | 0,03235994 | * |
| Misc | Water | 9,43E-05 | *** |
| Misc | Photo | 1,37E-06 | *** |